\documentclass[aps,prappl,twocolumn, superscriptaddress]{revtex4-1}
 \usepackage{graphicx}
\usepackage{physics}
\usepackage{xcolor}

\usepackage[breaklinks=true,colorlinks,citecolor=blue,linkcolor=blue,urlcolor=blue]{hyperref}

\draft 

\begin{document}


\title{Interplay between magnetism and superconductivity in a hybrid magnon-photon bilayer system} 



\author{Alberto Ghirri}
\email[mail to: ]{alberto.ghirri@nano.cnr.it}
\affiliation{Istituto Nanoscienze - CNR, Centro S3, via G. Campi 213/A, 41125, Modena, Italy.}
\author{Claudio Bonizzoni}
\affiliation{Dipartimento di Scienze Fisiche, Informatiche e Matematiche Universit\`a di Modena e Reggio Emilia, via G. Campi 213/A, 41125, Modena, Italy}
\affiliation{Istituto Nanoscienze - CNR, Centro S3, via G. Campi 213/A, 41125, Modena, Italy.}
\author{Maksut Maksutoglu}
\affiliation{Institute of Nanotechnology, Gebze Technical University, 41400, Gebze, Kocaeli, Turkey}
\author{Marco Affronte}
\affiliation{Dipartimento di Scienze Fisiche, Informatiche e Matematiche Universit\`a di Modena e Reggio Emilia, via G. Campi 213/A, 41125, Modena, Italy}
\affiliation{Istituto Nanoscienze - CNR, Centro S3, via G. Campi 213/A, 41125, Modena, Italy.}

\date{\today}

\begin{abstract}

Spin waves in magnetic films are affected by the vicinity to a superconductor. Here we focus on a bilayer stack made of an insulating Yttrium Iron Garnet (YIG) film and a high-$T_c$ YBCO superconducting planar resonator and report microwave transmission spectra to monitor the temperature evolution of magnon-photon polaritons. We show that the observed temperature dependence of normal mode splitting and frequency shift with respect to the unperturbed magnon mode can be ultimately related to the penetration depth of YBCO, as an effect of the interplay between spin waves and Meissner currents.

\end{abstract}

\maketitle 

The interplay between magnetism and superconductivity comprises a variety of physical phenomena observed when superconducting and magnetic layers get close one to another: the diamagnetism of an adjacent superconductor perturbs the propagation of spin waves in ferro(i)magnets \cite{GolovchanskiyJAP18, YuPRL22, BorstScience23, ZhouPRB23} and, in turns, the superconducting state of a film is affected by the profile of spin excitations in a neighbouring magnetic layer \cite{DobrovolskiyNatPhys19, IanovskaiaPRB23}. The description of these effects is intriguing and, just to start with, it will be useful to assess to which extent the Landau-Lifshitz-Gilbert (LLG) description of spin waves combined with the two-fluid model for the superconductor are effective to account for, at least, the non-dissipative effects of this interplay. 

High-$T_c$ superconducting resonators, in particular those based on $\mathrm{YBa_2Cu_3O_7}$ (YBCO) films, display critical temperature ($T_c$) above the liquid nitrogen boiling point, resilience in applied magnetic field and low damping of microwaves \cite{Hein, GhirriAPL15}. These characteristics have been successfully exploited for the manipulation of spin systems \cite{BonizzoniNPJ20, ArtziJMR21, BonizzoniApplMagnReson23, BonizzoniNPJQuant24} and for the implementation of hybrid spin-photon modes \cite{GhirriAPL15, GhirriPRA16, BonizzoniScieRep17, Velluire-PellatScieRep23}. Insulating and ferrimagnetic Yttrium Iron Garnet (YIG) is widely used in (planar) magnonic devices for its exceptionally low damping of magnetization precession \cite{GuerevichMelkov, PirroNatRevMater21}.

\begin{figure}[t]
\centering
\includegraphics[width=\linewidth] {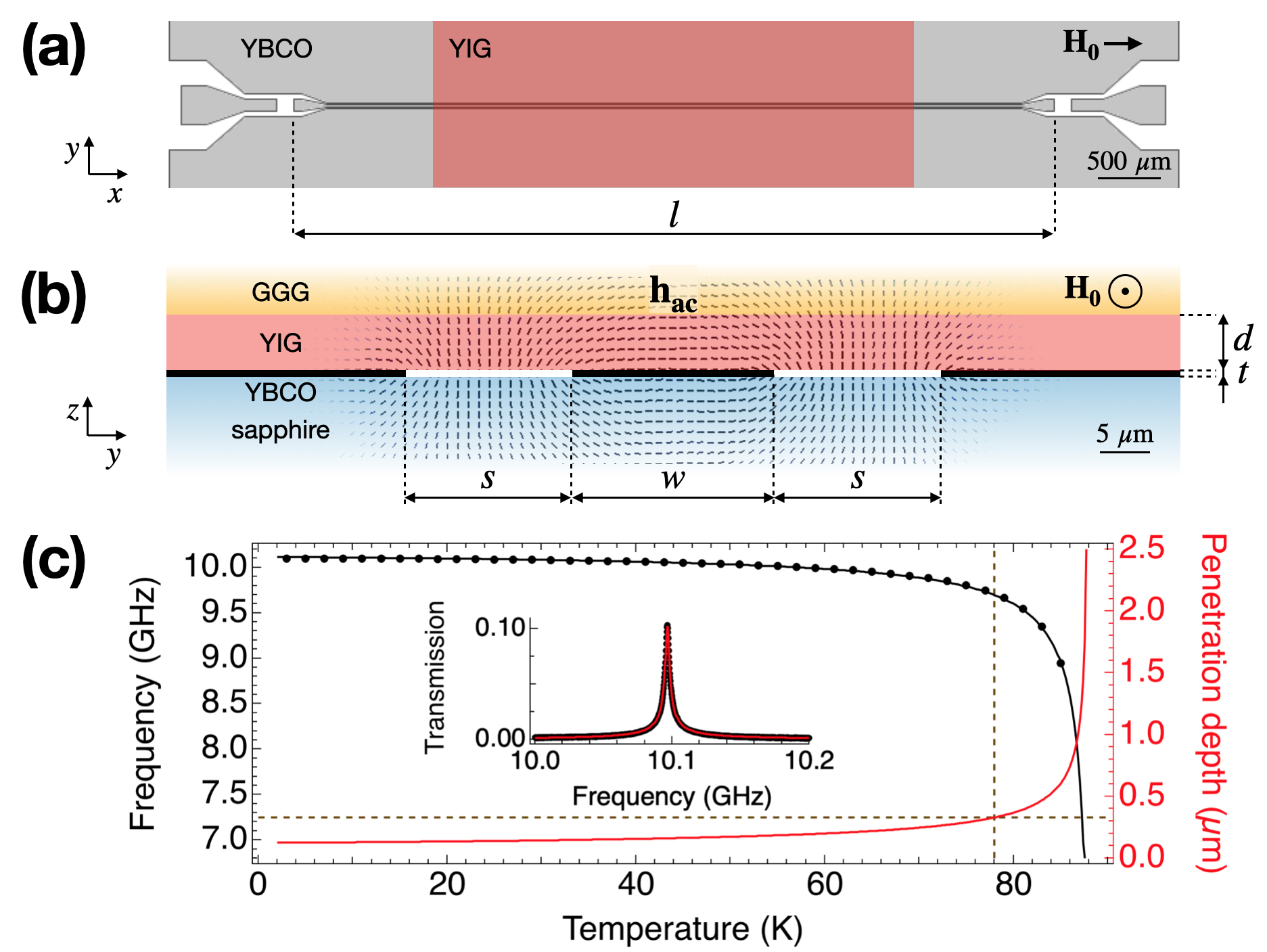}
\caption{(a) Top view and (b) vertical section of the YBCO resonator with the YIG film stacked on top. (c) Temperature dependence of the fundamental frequency ($\omega_c(T)$) of the bare resonator (circles); the black line is calculated with Eq.~\ref{eq:omega0_temp}. The penetration depth derived at the frequency of the resonator (Eq.~\ref{eq:penetration_depth}) is displayed in red. The horizontal dashed line indicates the thickness of the YBCO film ($t=330$~nm). Inset. $S_{21}(\omega)$ spectrum taken at 10~K (circles) and fit \cite{supplementary}. The incident power is $-15~\mathrm{dBm}$ in all the measurements.}
\label{fig:device}
\end{figure}

Magnetic imaging experiments have shown that underneath a superconducting layer the frequency of the spin waves is shifted, giving rise to hybrid spin-wave-Meissner-current modes \cite{BorstScience23}. Superconducting gates, which have been predicted to induce effects related also to the confinement \cite{KharlanArxiv23} and transport \cite{ZhouArxiv24} of magnons, have potential exploitation in novel magnonic devices \cite{YuPRL22, BorstScience23}. The ability to control the propagation of collective spin wave excitations in magnetic films offers indeed novel paradigms for data processing \cite{ChumakIEEE22}. In cavity magnonics, the strong coupling between magnon and photon modes has been observed using either magnetic crystals or films and proposed for exploitation in quantum technologies \cite{ Lachance-QuirionApplPhysExp19, RameshtiPhysRep22}. Superconducting circuits operated at microwave frequencies have been developed to investigate the hybridization of magnons and photons \cite{HueblPRL13, MorrisScieRep17, HouPRL19, LiPRL19, MandalAPL20, HaygoodPRAppl21, BaityAPL21, LiPRL22, BottcherArxiv23}, showing that superconductor-ferromagnet trilayer \cite{GolovchanskiyScAdv21, GolovchanskiyPRAppl21, SilaevPRB23} and bilayer \cite{GhirriPrAppl23} stacked structures can be exploited to achieve the ultrastrong coupling regime \cite{Kockum2019}. The possibility to span the whole temperature range of YBCO below $T_c$ is desirable both for testing the validity of theoretical models and for devising magnonic devices. Considering all these aspects, stacks of YIG/YBCO films represent an interesting case study for both fundamental and applied point of view.

Here we report on microwave transmission measurements carried out on a YIG/YBCO bilayer at different temperatures. Spectra acquired by using a broadband YBCO coplanar waveguide (CPW) above and below $T_c$ allow us to get a first characterizazion of the magnon spectrum in the frequency range of interest. Secondly, transmission spectra acquired with the superconducting resonator at decreasing temperatures show that the increase of the magnon-photon coupling is accompanied by a progressive shift of the anticrossing. We reproduce the trends observed in the experiments with a simple model that includes the interplay between spin wave excitations and Meissner currents, demonstrating that the evolution of the hybrid magnon-photon system can be directly related to the penetration depth of the superconductor.

CPW transmission lines and resonators were fabricated by optical lithography starting from superconducting YBCO films (thickness $t=330$~nm) deposited on sapphire \cite{supplementary}. The central conductor had width $w=(17 \pm 1)~\mu$m and separation $s=(14 \pm 1)~\mu$m from the lateral ground planes; the length was $l=6$~mm (Fig.~\ref{fig:device}(a,b)). Electromagnetic simulations \cite{GhirriPrAppl23} show that the microwave field is confined within a few tens of $\mu \mathrm{m}$ above the YBCO surface (Fig.~\ref{fig:device}(b)).
The transmission ($S_{21}$) spectrum of the bare resonator shows a half-wavelength fundamental mode whose frequency is $\omega_c(0)/2\pi  = 10.1~\mathrm{GHz}$ at the chosen reference temperature $T_0=10$~K (Inset in Fig.~\ref{fig:device}(c)). The fundamental mode frequency progressively decreases with increasing temperature up to $T_c$ (Fig.~\ref{fig:device}(c)). 

From the standard theory of distributed element transmission lines, the frequency of the resonator results in $\omega_c/2\pi=1/(2 l \sqrt{L C})$, where $C$ and $L=L_g+L_k$ are respectively the capacitance and the inductance per unit length, the latter given by the sum of the geometric ($L_g$) and kinetic ($L_k$) contributions. Being $C$ weakly depending on the temperature, the $\omega_c(T)$ dependence can be accounted to the change in the inductance \cite{VendikIEEE98, GhigoSuperSciTech04}
\begin{equation}
\omega_c \approx \omega_c(0) \sqrt{\frac{L[\lambda_L(T_0)]}{L[\lambda_L(T)]}},
\label{eq:omega0_temp}
\end{equation}
where $L$ depends, through $L_k$, upon the penetration depth of the superconductor. In the simplest two-fluid binomial approximation, the latter can be expressed as \cite{ProzorovSuperScieTech06}
\begin{eqnarray}
   \lambda_L=\frac{\lambda_L(0)}{\sqrt{1-\left (\frac{T}{T_c} \right)^{p}}},
   \label{eq:penetration_depth}
\end{eqnarray}
where $\lambda_L(0)$, is the London penetration length in the zero-temperature limit and $p=4/3$ for d-wave superconductors \cite{ProzorovSuperScieTech06, VendikIEEE98, GhigoSuperSciTech04}. The explicit relation to derive $\omega_c$ from Eqs.~\ref{eq:omega0_temp} and \ref{eq:penetration_depth} is reported in \cite{supplementary}. The best fit of the experimental data in Fig.~\ref{fig:device}(c) gives $\lambda_L(0)=124$~nm and $T_c=88.0$~K. These values well match the typical ones reported for high quality YBCO \cite{GhigoSuperSciTech04}.

\begin{figure}[t]
\centering
\includegraphics[width=\linewidth]{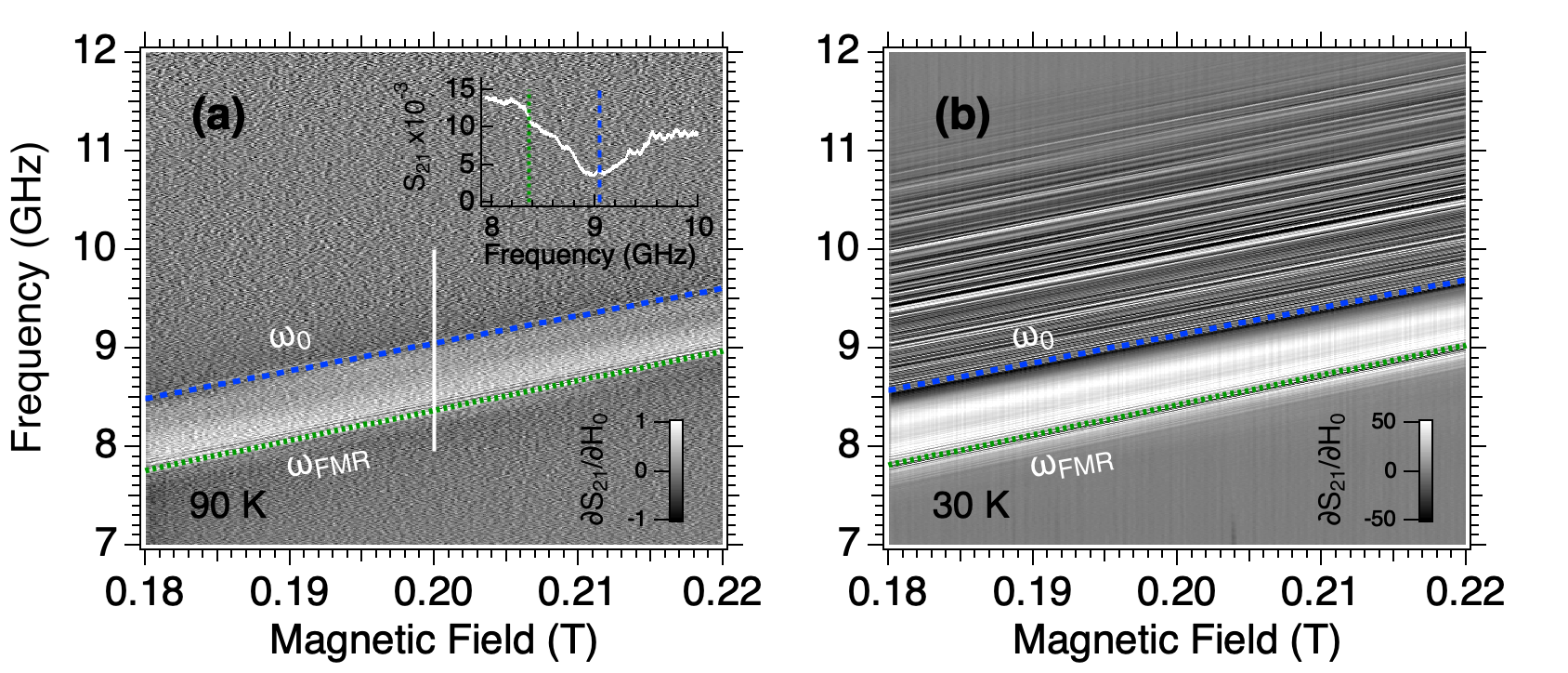}
\caption{Broadband ferromagnetic resonance spectra taken (a) above (90~K) and (b) well below (30~K) the critical temperature of YBCO. Transmission data are plotted as derivative with respect to the magnetic field, $\partial S_{21}/\partial H_0$. Green dashed lines display $\omega_{FMR}$, while blue dashed lines $\omega_{0}$ (Eq.~\ref{eq:KS_film_p0}). Inset. $S_{21}$ spectrum acquired at 0.2~T.}
\label{fig:broadband}
\end{figure}

\begin{figure*}[t]
\centering
\includegraphics[width=\linewidth]{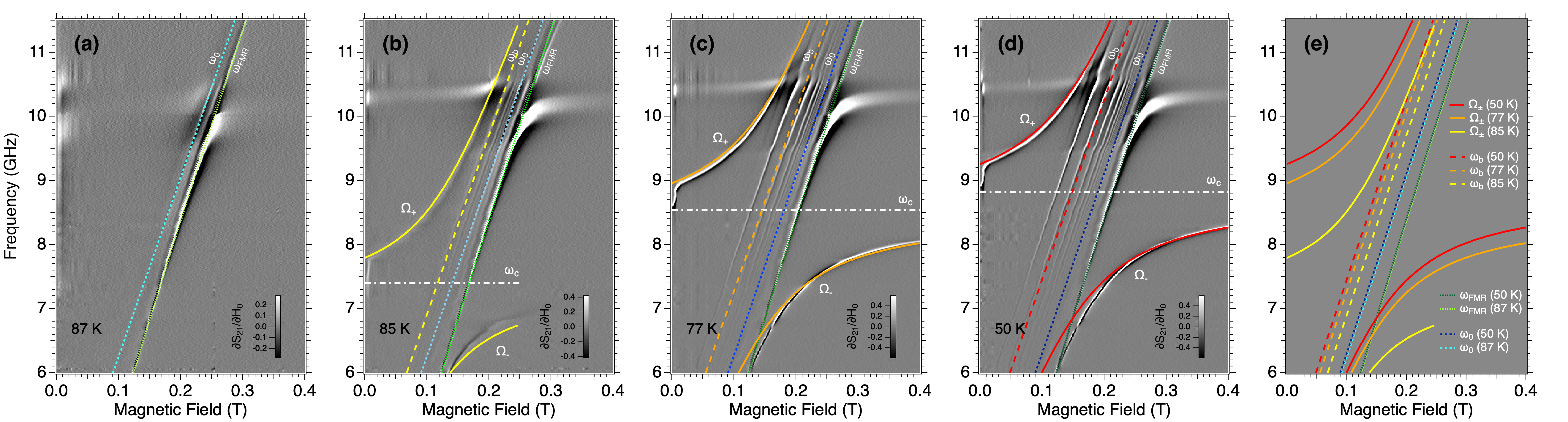}
\caption{(a-d) Derivative $\partial S_{21}/\partial H_0$ data showing the temperature evolution of the coupled spin wave and resonator modes (dataset B). The dash-dot lines indicate $\omega_c$ at each temperature. The dashed lines show $\omega_{FMR}$ (green), $\omega_{0}$ (blue) and the fit of the polaritonic modes $\Omega_{\pm}$ with Eq.~\ref{eq:polariton_freq}. (e) Direct comparison between the mode frequencies calculated at different temperatures.}
\label{fig:YIG_linear_data}
\end{figure*}

A YIG film with thickness $d=5~\mu$m grown on Gadolinium Gallium Garnet substrate (YIG/GGG) was positioned in contact with the YBCO device \cite{supplementary}. The external magnetic field was applied in the plane of the superconductor along the direction of the CPW line, $\mathbf{H_0}=H_0 \hat{x}$ (Fig.~\ref{fig:device}(a,b)). We first carried out transmission measurements by means of a YBCO broadband CPW line to investigate the magnon spectrum excited by the geometry of this type of antenna. Two spectral maps obtained above and below the observed transition temperature are shown in Fig.~\ref{fig:broadband}. Spectrum below $T_c$ (Fig.~\ref{fig:broadband}(b)) is much richer and shows modes that extend at high frequencies besides a set of resonances that are present also above $T_c$.
In the transmission spectrum we can identify an absorption band with characteristic frequencies (Fig.~\ref{fig:broadband}(a)). The lower limit of the band is given by the Kittel - or ferromagnetic resonance - frequency \cite{KittelPR48},  $\omega_{FMR}/2\pi=\gamma\mu_0\sqrt{H_0(H_0+M_s)}$, being $\mu_0=4 \pi \times 10^{-7}$~H/m the vacuum permeability and $\gamma=28.02~\mathrm{GHz/T}$ the electron's gyromagnetic ratio. The saturation magnetization of YIG, $M_s$, is assumed to vary with temperature \cite{MaierFlaigPRB17} between $\mu_0M_s (\mathrm{10~K})=0.246$~T and $\mu_0M_s(\mathrm{90~K})=0.239$~T \cite{supplementary}. 

Besides, we note that maximum absorption coincides with the frequency $\omega_0 > \omega_{FMR}$ (Inset in Fig.~\ref{fig:broadband}(a)). CPW lines with oscillating field $\mathbf{h_{ac}} \perp \hat{x}$ can excite spin wave modes having wavenumber $k_y \approx 2 \pi/s$ \cite{MaksymovPhysE15}. According to the seminal work of Kalinikos and Slavin (KS) for a YIG film \cite{KalinikosJPhysC86}, in this geometry the spectrum of spin wave modes consists of the lowest dipole-dominated mode characterized by a quasi-uniform thickness profile and higher exchange-dominated spin wave resonance (SWR) modes \cite{supplementary}. The lowest mode can be described by:  
\begin{align}
    \begin{split}
    & \omega_0 / 2 \pi= \\ & = \mu_0 \gamma \sqrt{H_0(H_0+M_s)+(M_s)^2 P_{00}(k_yd) \left(1-P_{00}(k_yd)\right)}, 
    \label{eq:KS_film_p0}
    \end{split}
\end{align}
that for $k_y d \ll 1$ is similar to the Damon-Eshbach expression \cite{DemokritovSpringer21, GhirriPrAppl23}. In Eq.~\ref{eq:KS_film_p0}, $k_y$ is the wavevector of the spin wave mode and $P_{00}=1+[(1-\exp(-k_y d))/k_y d]$ \cite{DemokritovSpringer21}. As shown in Fig.~\ref{fig:broadband}(a), $\omega_0$ can reproduce the position of the transmission minimum as a function of the magnetic field. From the direct comparison between Eq.~\ref{eq:KS_film_p0} and the experimental data, we obtain $k_y=3 \times 10^5~\mathrm{rad~m}^{-1}$, consistently with our geometrical factors, i.e. $k_y \approx 2 \pi/s$ \cite{MaksymovPhysE15}.

As the temperature decreases below $T_c$ (Fig.~\ref{fig:broadband}(b)), $\omega_0$ is nearly temperature independent. By using $k_y=3 \times 10^5~\mathrm{rad~m}^{-1}$ we can be reproduce $\omega_0(H_0)$ in the whole 10-90~K temperature range, because Eq.~\ref{eq:KS_film_p0} weakly depends from temperature through $M_s$ \cite{supplementary}. Yet, the main differences between panel (a) and (b) are related to the presence of an absorption band that progressively widens below $T_c$, showing a progressively increasing number of spin wave resonance modes at frequencies $\omega>\omega_0$ as a result of the occurrence of superconductivity in the YBCO layer. However, the complexity of the experimental spectra maps makes it difficult to the discern between unperturbed features of the YIG film and those due to the interplay with the superconductor \cite{supplementary}.

The spectral maps show more distinguishable features when the same YIG film is positioned on the CPW resonator (Fig.~\ref{fig:YIG_linear_data}). Note that, in this second set of experiments, the CPW resonator has the same lateral dimensions as the broadband CPW line used in the previous experiments. We can recognise a clear trend of the spectrum as temperature decreases. At $T=87$~K the resonator mode is not visible but weak absorption lines can be noticed due to the magnetic film. In Fig.~\ref{fig:YIG_linear_data}(a) we can identify the Kittel mode, $\omega_{FMR}$, and the quasi-uniform thickness mode, $\omega_0$. Consistently with data in Fig.~\ref{fig:broadband}, the latter can be reproduced by Eq.~\ref{eq:KS_film_p0} with $k_y=3 \times 10^5~\mathrm{rad~m}^{-1}$. At $T=85$~K, the resonator mode appears with frequency $\omega_c/2\pi \approx 7.4$~GHz (Fig.~\ref{fig:YIG_linear_data}(b)). Due to the permittivity of the YIG/GGG sample, this value is lower than what was obtained at the same temperature using the bare resonator. The spectral map here shows  two polaritonic branches as a result of the hybridization between magnons and photons. Below 77~K, the fundamental mode frequency increases and the splitting (2$g$) between the two branches progressively increases up to the maximum value at the lowest temperature (Fig.~\ref{fig:YIG_linear_data}(c,d)). This temperature-dependent widening of the anticrossing gap is determined by the progressive displacement of the polariton branches towards higher frequencies and by the shift of the upper polariton towards lower magnetic fields. As shown in \cite{supplementary}, these trends have been reproduced in another dataset and obtained also with a different resonator. They can be ascribed to the occurrence of the superconductivity in the YBCO layer \cite{GolovchanskiyPRAppl21, GolovchanskiyPRAppl23, GhirriPrAppl23}, as quantified in the following.

We model the physical system comprising of resonator modes interacting with collective spin wave excitations on the basis of a modified Hopfield Hamiltonian, whose eigenvalues reads \cite{GhirriPrAppl23}
\begin{equation}
    \label{eq:polariton_freq}
    \Omega_\pm = \frac{1}{\sqrt{2}} \sqrt{\tilde{\omega}_{c}^2 + \omega_{b}^2 \pm \sqrt{\left(\tilde{\omega}_{c}^2 - \omega_{b}^2 \right)^2 + 16 \omega_{c} \omega_{b} g^2}} ~,
\end{equation}
where $\tilde{\omega}_{c} = \sqrt{\omega_{c} (\omega_{c} + 4 \beta)}$, being $\beta$ the diamagnetic term that, however, we take vanishingly small. In Eq.~\ref{eq:polariton_freq} we consider a single magnetic mode coupled to the resonator \cite{GhirriPrAppl23}, whose frequency is 
\begin{equation}
    \label{eq:spin-wave-Meissner-current_freq}
    \omega_b=\omega_{0}+\delta_{sc}, 
\end{equation}
being $\omega_{0}$ the frequency of the lowest YIG mode (Eq.~\ref{eq:KS_film_p0}) and $\delta_{sc}$ the temperature-dependent shift. $\omega_b$ corresponds to the frequency of the effective magnon mode that best couple with the superconducting resonator. Neglecting dissipative effects in the superconductor in first instance, the frequency shift $\delta_{sc}$ can be quantified by self-consistently including the spin-wave induced Meissner currents in the Landau-Lifshitz-Gilbert equation to obtain \cite{BorstScience23}
\begin{equation}
    \delta_{sc} \approx \gamma \mu_0 M_s k_y d r \frac{1-e^{-\frac{2t}{\lambda_L}}}{(k_y \lambda_L+1)^2-(k_y \lambda_L-1)^2 e^{-\frac{2t}{\lambda_L}}},
    \label{eq:delta}
\end{equation} 
where $r$ is a dimensionless geometrical factor.

The polaritonic branches (Eq.~\ref{eq:polariton_freq}) depend on temperature also through the frequency of the resonator (Eq.~\ref{eq:omega0_temp}). Additionally, the spin-photon coupling is \cite{TosiAIPAdv14}
\begin{equation}
    g_{s} = \frac{1}{4}\gamma b_{vac}\approx \frac{\mu_0 \omega_{c}}{4 w} \sqrt{\frac{h}{Z_0}},
\label{eq:gs}
\end{equation}
being $b_{vac}$ the vacuum magnetic field of the resonator and $Z_0=58~\mathrm{\Omega}$ is the nominal impedance of the CPW line. Eq.~\ref{eq:gs} also depends from the temperature through $\omega_c$.

We used Eq.~\ref{eq:polariton_freq} to reproduce the magnetic field dispersion of the polaritonic branches (dataset A in \cite{supplementary} and dataset B in Fig.~\ref{fig:YIG_linear_data}) and to fix key values of the problem. We considered $\omega_{c}$ and $\delta_{sc}$ free parameters, whereas the collective coupling strength $g=g_{s} \sqrt{2 s_{\mathrm{Fe}} N_s}$, being $s_{\mathrm{Fe}}=5/2$ the single-ion spin of Fe$^{3+}$, was calculated using the number of spins ($N_s$), which was assumed to be temperature-independent for each dataset. Following this approach, we reproduced the evolution of the polaritons in Fig.~\ref{fig:YIG_linear_data}, as well as in the additional spectral maps shown in \cite{supplementary}. Panel (e) summarizes the temperature evolution of the relevant modes: as the temperature is lowered we note that $\omega_b$ shifts towards lower magnetic fields much faster than $\omega_0$ and $\omega_{FMR}$.  Additional modes visible in panels (a-d) can be attributed to the coupling between spin waves and the higher mode of the resonator at $\approx 10.5~\mathrm{GHz}$, as discussed more in detail in \cite{supplementary}.
\begin{figure}[h]
\centering
\includegraphics[width=0.9\linewidth]{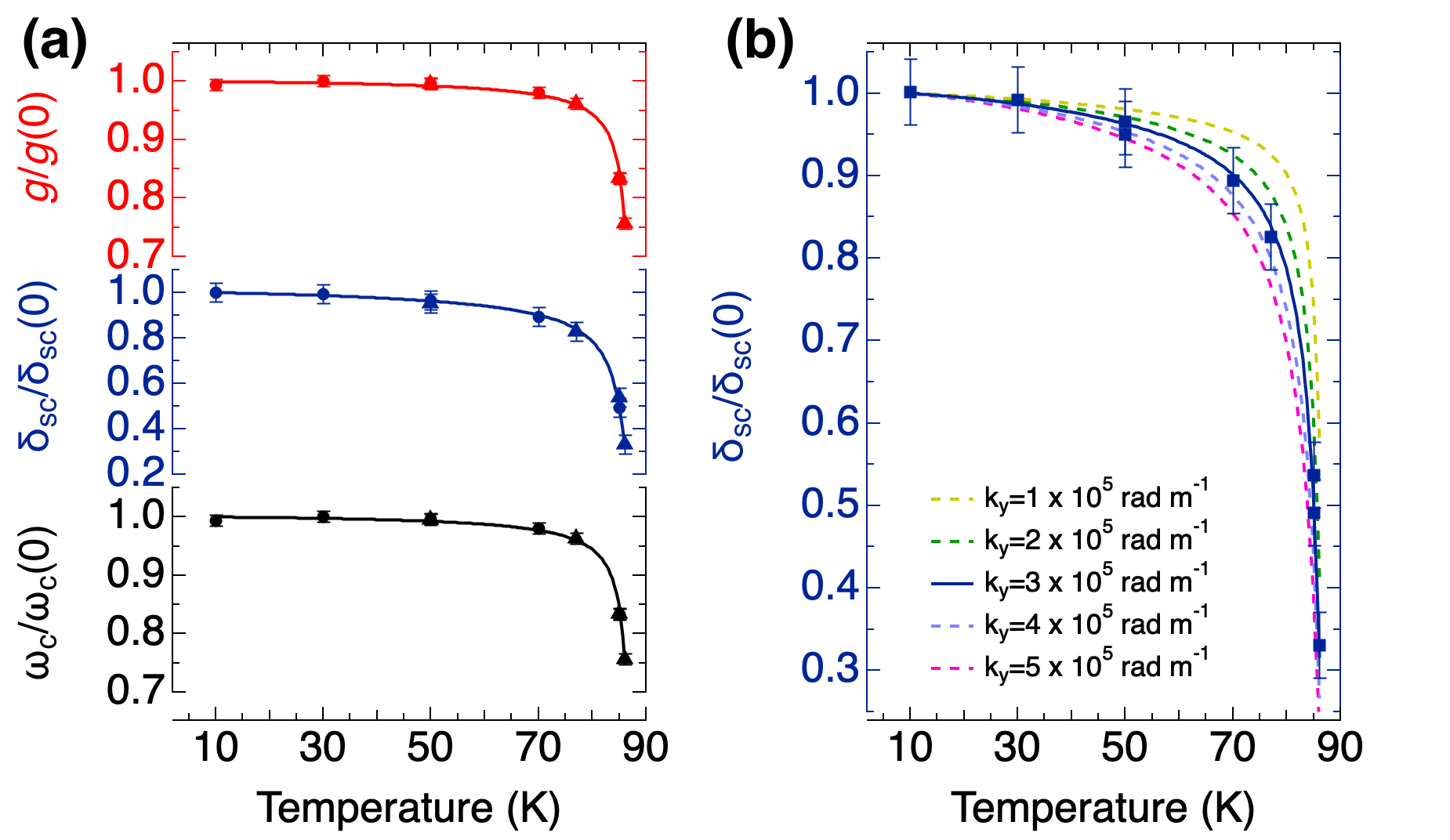}
\caption{(a) Temperature dependence of the parameters obtained from the fit. The circle and triangle symbols indicate the parameters extracted respectively for datasets A and B;  solid lines show the calculated curves. The vertical scale is normalized to $\omega_c^A(0)/2\pi=9.19$~GHz, $g^A(0)/2\pi=1.72$~GHz and $\delta_{sc}^A(0)/2\pi=1.12$~GHz for dataset A, $\omega_c^B(0)/2\pi=8.86$~GHz, $g^B(0)/2\pi=1.77$~GHz and $\delta_{sc}^B(0)/2\pi=1.21$~GHz for dataset B. The number of spins resulted in $N_s^A=0.93 \times 10^{15}$ and $N_s^B=1.05 \times 10^{15}$. (b) Comparison between the normalized frequency shift and the curves obtained from Eq.~\ref{eq:delta} with different wavenumber $k_y$.}
\label{fig:YIG_linear_param}
\end{figure}

We now turn our attention to the temperature dependence of the parameters $\omega_c$, $g$ and $\delta_{sc}$ obtained from the comparison between Eq.~\ref{eq:polariton_freq} and the experimental spectra (Fig.~\ref{fig:YIG_linear_param}(a)). We used Eq.~\ref{eq:omega0_temp} to fit the temperature dependence of $\omega_{c}$: $\lambda_L(0)$ and $p$ were kept fixed to the values determined for the bare resonator while $T_c=86.9~\mathrm{K}$ is obtained. Furthermore, the collective coupling strength ($g$) can be easily reproduced using Eq.~\ref{eq:gs} along with the fitted values of $\omega_{c}$ and the number of spins $N_s \approx 1 \times 10^{15}$ (Fig.~\ref{fig:YIG_linear_param}(a)). Note that the maximum value of the collective coupling $g(0)/2\pi \approx 1.7~\mathrm{GHz}> \omega_c(0)/10$ confirms that the ultrastrong coupling regime is achieved \cite{Kockum2019, GhirriPrAppl23}.

Interestingly, the evolution of the frequency shift parameter $\delta_{sc}$, shows a well-defined temperature dependence: $\delta_{sc}$ is small for $T \lesssim T_c$, conversely for $T \ll T_c$ the frequency shift saturates to the maximum value $\delta_{sc}(0)/2\pi \approx 1.1-1.2$~GHz (Fig.~\ref{fig:YIG_linear_param}(a)). In the two-fluid model of the superconductor, the dissipation due to normal electrons is related to the real part of the complex conductivity $\sigma (\omega)=\sigma_1(\omega)-i \sigma_2(\omega)$. For YBCO films with thickness on the order of few hundred nanometers, the expected $\sigma_1/\sigma_2$ ratio is of the order of one tenth or less just below $T_c$ \cite{KrupkaIEEE13, supplementary} thus supporting the validity of the non-dissipative approximation. With Eq.~\ref{eq:delta} we calculate the temperature dependence of $\delta_{sc}$ by using the penetration depth estimated with Eq.~$\ref{eq:penetration_depth}$ and the previously reported values of $\lambda_L(0)$, $T_c$ and $p$. Additionally, we use $k_y=3 \times 10^5~\mathrm{rad~m}^{-1}$ as obtained from the fit of the spectra in Fig.~\ref{fig:broadband} and \ref{fig:YIG_linear_data}. Note that the temperature dependence of Eq.~\ref{eq:delta} mainly derives from the temperature dependence of the penetration depth (Eq.~\ref{eq:penetration_depth}) and, only partially from the weak temperature dependence of the saturation magnetization in this range \cite{supplementary}. 

Fig.~\ref{fig:YIG_linear_param}(a) shows that Eq.~\ref{eq:delta} successfully reproduce the temperature dependence of the fitted values of $\delta_{sc}$. To further test the comparison between Eq.~\ref{eq:delta} and the values of $\delta_{sc}$ derived from the fit, in Fig.~\ref{fig:YIG_linear_param}(b) we plot the frequency shift calculated for different wavenumbers $k_y$. The excellent match with $k_y=3 \times 10^5~\mathrm{rad~m}^{-1}$ confirms the consistency of our results. The validity of this analysis is supported also by the results obtained with a YBCO resonator having different geometrical parameters \cite{supplementary}.

To summarize, we have studied the temperature dependence of transmission spectra acquired on a YIG/YBCO bilayer system. Below $T_c$ and with decreasing temperatures, the magnon-photon polaritons are progressively shifted due, primarily, to the increase of the resonator frequency that determines also the increase of the coupling strength. Furthermore, we have observed that the anticrossing is not centered on the unperturbed mode of the YIG film, but is progressively displaced up to more than 1~GHz at low temperature. Both effects are essentially governed by the temperature dependence of the penetration depth, the latter in particular follows the trend expected from the interplay between spin waves and Meissner currents.

The excellent fit of the simulated curves to the experimental behavior confirms the efficiency of the two-fluid model to account for the non-dissipative interplay between magnetic excitations and an adjacent superconducting layer. Line broadening or microscopic visualization of the mixed state of the superconductor will certainly require to consider dissipative effects that can be included in the LLG equation at least in a phenomenological way \cite{BorstScience23}. Considering the recent proposals to exploit the diamagnetism of superconductors to gate the wave propagation in planar magnonic devices \cite{BorstScience23, YuPRL22, ZhouArxiv24}, and our results that demonstrate the achievement of large coupling strengths and shifts of the magnon mode even at liquid nitrogen temperatures, we foresee the possibility to use high-$T_c$ superconductors in the context of superconducting magnonics.

\begin{acknowledgments}
This work was partially supported by the European Community through FET Open SUPERGALAX project (grant agreement No.~863313); by NATO Science
for Peace and Security Programme (NATO SPS Project No.~G5859) and by US. Office of Naval Research award N62909-23-1-2079.
\end{acknowledgments}

\bibliography{biblio}

\begin{thebibliography}{54}%
\makeatletter
\providecommand \@ifxundefined [1]{%
 \@ifx{#1\undefined}
}%
\providecommand \@ifnum [1]{%
 \ifnum #1\expandafter \@firstoftwo
 \else \expandafter \@secondoftwo
 \fi
}%
\providecommand \@ifx [1]{%
 \ifx #1\expandafter \@firstoftwo
 \else \expandafter \@secondoftwo
 \fi
}%
\providecommand \natexlab [1]{#1}%
\providecommand \enquote  [1]{``#1''}%
\providecommand \bibnamefont  [1]{#1}%
\providecommand \bibfnamefont [1]{#1}%
\providecommand \citenamefont [1]{#1}%
\providecommand \href@noop [0]{\@secondoftwo}%
\providecommand \href [0]{\begingroup \@sanitize@url \@href}%
\providecommand \@href[1]{\@@startlink{#1}\@@href}%
\providecommand \@@href[1]{\endgroup#1\@@endlink}%
\providecommand \@sanitize@url [0]{\catcode `\\12\catcode `\$12\catcode
  `\&12\catcode `\#12\catcode `\^12\catcode `\_12\catcode `\%12\relax}%
\providecommand \@@startlink[1]{}%
\providecommand \@@endlink[0]{}%
\providecommand \url  [0]{\begingroup\@sanitize@url \@url }%
\providecommand \@url [1]{\endgroup\@href {#1}{\urlprefix }}%
\providecommand \urlprefix  [0]{URL }%
\providecommand \Eprint [0]{\href }%
\providecommand \doibase [0]{http://dx.doi.org/}%
\providecommand \selectlanguage [0]{\@gobble}%
\providecommand \bibinfo  [0]{\@secondoftwo}%
\providecommand \bibfield  [0]{\@secondoftwo}%
\providecommand \translation [1]{[#1]}%
\providecommand \BibitemOpen [0]{}%
\providecommand \bibitemStop [0]{}%
\providecommand \bibitemNoStop [0]{.\EOS\space}%
\providecommand \EOS [0]{\spacefactor3000\relax}%
\providecommand \BibitemShut  [1]{\csname bibitem#1\endcsname}%
\let\auto@bib@innerbib\@empty
\bibitem [{\citenamefont {Golovchanskiy}\ \emph {et~al.}(2018)\citenamefont
  {Golovchanskiy}, \citenamefont {Abramov}, \citenamefont {Stolyarov},
  \citenamefont {Ryazanov}, \citenamefont {Golubov},\ and\ \citenamefont
  {Ustinov}}]{GolovchanskiyJAP18}%
  \BibitemOpen
  \bibfield  {author} {\bibinfo {author} {\bibfnamefont {I.~A.}\ \bibnamefont
  {Golovchanskiy}}, \bibinfo {author} {\bibfnamefont {N.~N.}\ \bibnamefont
  {Abramov}}, \bibinfo {author} {\bibfnamefont {V.~S.}\ \bibnamefont
  {Stolyarov}}, \bibinfo {author} {\bibfnamefont {V.~V.}\ \bibnamefont
  {Ryazanov}}, \bibinfo {author} {\bibfnamefont {A.~A.}\ \bibnamefont
  {Golubov}}, \ and\ \bibinfo {author} {\bibfnamefont {A.~V.}\ \bibnamefont
  {Ustinov}},\ }\href {\doibase 10.1063/1.5077086} {\bibfield  {journal}
  {\bibinfo  {journal} {Journal of Applied Physics}\ }\textbf {\bibinfo
  {volume} {124}},\ \bibinfo {pages} {233903} (\bibinfo {year}
  {2018})}\BibitemShut {NoStop}%
\bibitem [{\citenamefont {Yu}\ and\ \citenamefont {Bauer}(2022)}]{YuPRL22}%
  \BibitemOpen
  \bibfield  {author} {\bibinfo {author} {\bibfnamefont {T.}~\bibnamefont
  {Yu}}\ and\ \bibinfo {author} {\bibfnamefont {G.~E.~W.}\ \bibnamefont
  {Bauer}},\ }\href {\doibase 10.1103/PhysRevLett.129.117201} {\bibfield
  {journal} {\bibinfo  {journal} {Phys. Rev. Lett.}\ }\textbf {\bibinfo
  {volume} {129}},\ \bibinfo {pages} {117201} (\bibinfo {year}
  {2022})}\BibitemShut {NoStop}%
\bibitem [{\citenamefont {Borst}\ \emph {et~al.}(2023)\citenamefont {Borst},
  \citenamefont {Vree}, \citenamefont {Lowther}, \citenamefont {Teepe},
  \citenamefont {Kurdi}, \citenamefont {Bertelli}, \citenamefont {Simon},
  \citenamefont {Blanter},\ and\ \citenamefont {van~der Sar}}]{BorstScience23}%
  \BibitemOpen
  \bibfield  {author} {\bibinfo {author} {\bibfnamefont {M.}~\bibnamefont
  {Borst}}, \bibinfo {author} {\bibfnamefont {P.~H.}\ \bibnamefont {Vree}},
  \bibinfo {author} {\bibfnamefont {A.}~\bibnamefont {Lowther}}, \bibinfo
  {author} {\bibfnamefont {A.}~\bibnamefont {Teepe}}, \bibinfo {author}
  {\bibfnamefont {S.}~\bibnamefont {Kurdi}}, \bibinfo {author} {\bibfnamefont
  {I.}~\bibnamefont {Bertelli}}, \bibinfo {author} {\bibfnamefont {B.~G.}\
  \bibnamefont {Simon}}, \bibinfo {author} {\bibfnamefont {Y.~M.}\ \bibnamefont
  {Blanter}}, \ and\ \bibinfo {author} {\bibfnamefont {T.}~\bibnamefont
  {van~der Sar}},\ }\href {\doibase 10.1126/science.adj7576} {\bibfield
  {journal} {\bibinfo  {journal} {Science}\ }\textbf {\bibinfo {volume}
  {382}},\ \bibinfo {pages} {430} (\bibinfo {year} {2023})}\BibitemShut
  {NoStop}%
\bibitem [{\citenamefont {Zhou}\ and\ \citenamefont {Yu}(2023)}]{ZhouPRB23}%
  \BibitemOpen
  \bibfield  {author} {\bibinfo {author} {\bibfnamefont {X.-H.}\ \bibnamefont
  {Zhou}}\ and\ \bibinfo {author} {\bibfnamefont {T.}~\bibnamefont {Yu}},\
  }\href {\doibase 10.1103/PhysRevB.108.144405} {\bibfield  {journal} {\bibinfo
   {journal} {Phys. Rev. B}\ }\textbf {\bibinfo {volume} {108}},\ \bibinfo
  {pages} {144405} (\bibinfo {year} {2023})}\BibitemShut {NoStop}%
\bibitem [{\citenamefont {Dobrovolskiy}\ \emph {et~al.}(2019)\citenamefont
  {Dobrovolskiy}, \citenamefont {Sachser}, \citenamefont {Br{\"a}cher},
  \citenamefont {B{\"o}ttcher}, \citenamefont {Kruglyak}, \citenamefont {Vovk},
  \citenamefont {Shklovskij}, \citenamefont {Huth}, \citenamefont
  {Hillebrands},\ and\ \citenamefont {Chumak}}]{DobrovolskiyNatPhys19}%
  \BibitemOpen
  \bibfield  {author} {\bibinfo {author} {\bibfnamefont {O.~V.}\ \bibnamefont
  {Dobrovolskiy}}, \bibinfo {author} {\bibfnamefont {R.}~\bibnamefont
  {Sachser}}, \bibinfo {author} {\bibfnamefont {T.}~\bibnamefont
  {Br{\"a}cher}}, \bibinfo {author} {\bibfnamefont {T.}~\bibnamefont
  {B{\"o}ttcher}}, \bibinfo {author} {\bibfnamefont {V.~V.}\ \bibnamefont
  {Kruglyak}}, \bibinfo {author} {\bibfnamefont {R.~V.}\ \bibnamefont {Vovk}},
  \bibinfo {author} {\bibfnamefont {V.~A.}\ \bibnamefont {Shklovskij}},
  \bibinfo {author} {\bibfnamefont {M.}~\bibnamefont {Huth}}, \bibinfo {author}
  {\bibfnamefont {B.}~\bibnamefont {Hillebrands}}, \ and\ \bibinfo {author}
  {\bibfnamefont {A.~V.}\ \bibnamefont {Chumak}},\ }\href {\doibase
  10.1038/s41567-019-0428-5} {\bibfield  {journal} {\bibinfo  {journal} {Nature
  Physics}\ }\textbf {\bibinfo {volume} {15}},\ \bibinfo {pages} {477}
  (\bibinfo {year} {2019})}\BibitemShut {NoStop}%
\bibitem [{\citenamefont {Ianovskaia}\ \emph {et~al.}(2023)\citenamefont
  {Ianovskaia}, \citenamefont {Bobkov},\ and\ \citenamefont
  {Bobkova}}]{IanovskaiaPRB23}%
  \BibitemOpen
  \bibfield  {author} {\bibinfo {author} {\bibfnamefont {A.~S.}\ \bibnamefont
  {Ianovskaia}}, \bibinfo {author} {\bibfnamefont {A.~M.}\ \bibnamefont
  {Bobkov}}, \ and\ \bibinfo {author} {\bibfnamefont {I.~V.}\ \bibnamefont
  {Bobkova}},\ }\href {\doibase 10.1103/PhysRevB.108.214501} {\bibfield
  {journal} {\bibinfo  {journal} {Phys. Rev. B}\ }\textbf {\bibinfo {volume}
  {108}},\ \bibinfo {pages} {214501} (\bibinfo {year} {2023})}\BibitemShut
  {NoStop}%
\bibitem [{\citenamefont {Hein}(1999)}]{Hein}%
  \BibitemOpen
  \bibfield  {author} {\bibinfo {author} {\bibfnamefont {M.}~\bibnamefont
  {Hein}},\ }\href@noop {} {\emph {\bibinfo {title}
  {High-Temperature-Superconductor Thin Films at Microwave Frequencies}}}\
  (\bibinfo  {publisher} {Springer, Berlin},\ \bibinfo {year}
  {1999})\BibitemShut {NoStop}%
\bibitem [{\citenamefont {Ghirri}\ \emph {et~al.}(2015)\citenamefont {Ghirri},
  \citenamefont {Bonizzoni}, \citenamefont {Gerace}, \citenamefont {Sanna},
  \citenamefont {Cassinese},\ and\ \citenamefont {Affronte}}]{GhirriAPL15}%
  \BibitemOpen
  \bibfield  {author} {\bibinfo {author} {\bibfnamefont {A.}~\bibnamefont
  {Ghirri}}, \bibinfo {author} {\bibfnamefont {C.}~\bibnamefont {Bonizzoni}},
  \bibinfo {author} {\bibfnamefont {D.}~\bibnamefont {Gerace}}, \bibinfo
  {author} {\bibfnamefont {S.}~\bibnamefont {Sanna}}, \bibinfo {author}
  {\bibfnamefont {A.}~\bibnamefont {Cassinese}}, \ and\ \bibinfo {author}
  {\bibfnamefont {M.}~\bibnamefont {Affronte}},\ }\href {\doibase
  10.1063/1.4920930} {\bibfield  {journal} {\bibinfo  {journal} {Applied
  Physics Letters}\ }\textbf {\bibinfo {volume} {106}},\ \bibinfo {pages}
  {184101} (\bibinfo {year} {2015})}\BibitemShut {NoStop}%
\bibitem [{\citenamefont {Bonizzoni}\ \emph {et~al.}(2020)\citenamefont
  {Bonizzoni}, \citenamefont {Ghirri}, \citenamefont {Santanni}, \citenamefont
  {Atzori}, \citenamefont {Sorace}, \citenamefont {Sessoli},\ and\
  \citenamefont {Affronte}}]{BonizzoniNPJ20}%
  \BibitemOpen
  \bibfield  {author} {\bibinfo {author} {\bibfnamefont {C.}~\bibnamefont
  {Bonizzoni}}, \bibinfo {author} {\bibfnamefont {A.}~\bibnamefont {Ghirri}},
  \bibinfo {author} {\bibfnamefont {F.}~\bibnamefont {Santanni}}, \bibinfo
  {author} {\bibfnamefont {M.}~\bibnamefont {Atzori}}, \bibinfo {author}
  {\bibfnamefont {L.}~\bibnamefont {Sorace}}, \bibinfo {author} {\bibfnamefont
  {R.}~\bibnamefont {Sessoli}}, \ and\ \bibinfo {author} {\bibfnamefont
  {M.}~\bibnamefont {Affronte}},\ }\href {\doibase 10.1038/s41534-020-00296-9}
  {\bibfield  {journal} {\bibinfo  {journal} {npj Quantum Information}\
  }\textbf {\bibinfo {volume} {6}},\ \bibinfo {pages} {68} (\bibinfo {year}
  {2020})}\BibitemShut {NoStop}%
\bibitem [{\citenamefont {Artzi}\ \emph {et~al.}(2022)\citenamefont {Artzi},
  \citenamefont {Yishay}, \citenamefont {Fanciulli}, \citenamefont {Jbara},\
  and\ \citenamefont {Blank}}]{ArtziJMR21}%
  \BibitemOpen
  \bibfield  {author} {\bibinfo {author} {\bibfnamefont {Y.}~\bibnamefont
  {Artzi}}, \bibinfo {author} {\bibfnamefont {Y.}~\bibnamefont {Yishay}},
  \bibinfo {author} {\bibfnamefont {M.}~\bibnamefont {Fanciulli}}, \bibinfo
  {author} {\bibfnamefont {M.}~\bibnamefont {Jbara}}, \ and\ \bibinfo {author}
  {\bibfnamefont {A.}~\bibnamefont {Blank}},\ }\href {\doibase
  https://doi.org/10.1016/j.jmr.2021.107102} {\bibfield  {journal} {\bibinfo
  {journal} {Journal of Magnetic Resonance}\ }\textbf {\bibinfo {volume}
  {334}},\ \bibinfo {pages} {107102} (\bibinfo {year} {2022})}\BibitemShut
  {NoStop}%
\bibitem [{\citenamefont {Bonizzoni}\ \emph {et~al.}(2023)\citenamefont
  {Bonizzoni}, \citenamefont {Maksutoglu}, \citenamefont {Ghirri},
  \citenamefont {van Tol}, \citenamefont {Rameev},\ and\ \citenamefont
  {Affronte}}]{BonizzoniApplMagnReson23}%
  \BibitemOpen
  \bibfield  {author} {\bibinfo {author} {\bibfnamefont {C.}~\bibnamefont
  {Bonizzoni}}, \bibinfo {author} {\bibfnamefont {M.}~\bibnamefont
  {Maksutoglu}}, \bibinfo {author} {\bibfnamefont {A.}~\bibnamefont {Ghirri}},
  \bibinfo {author} {\bibfnamefont {J.}~\bibnamefont {van Tol}}, \bibinfo
  {author} {\bibfnamefont {B.}~\bibnamefont {Rameev}}, \ and\ \bibinfo {author}
  {\bibfnamefont {M.}~\bibnamefont {Affronte}},\ }\href {\doibase
  10.1007/s00723-022-01505-8} {\bibfield  {journal} {\bibinfo  {journal}
  {Applied Magnetic Resonance}\ }\textbf {\bibinfo {volume} {54}},\ \bibinfo
  {pages} {143} (\bibinfo {year} {2023})}\BibitemShut {NoStop}%
\bibitem [{\citenamefont {Bonizzoni}\ \emph {et~al.}(2024)\citenamefont
  {Bonizzoni}, \citenamefont {Ghirri}, \citenamefont {Santanni},\ and\
  \citenamefont {Affronte}}]{BonizzoniNPJQuant24}%
  \BibitemOpen
  \bibfield  {author} {\bibinfo {author} {\bibfnamefont {C.}~\bibnamefont
  {Bonizzoni}}, \bibinfo {author} {\bibfnamefont {A.}~\bibnamefont {Ghirri}},
  \bibinfo {author} {\bibfnamefont {F.}~\bibnamefont {Santanni}}, \ and\
  \bibinfo {author} {\bibfnamefont {M.}~\bibnamefont {Affronte}},\ }\href
  {\doibase 10.1038/s41534-024-00838-5} {\bibfield  {journal} {\bibinfo
  {journal} {npj Quantum Information}\ }\textbf {\bibinfo {volume} {10}},\
  \bibinfo {pages} {41} (\bibinfo {year} {2024})}\BibitemShut {NoStop}%
\bibitem [{\citenamefont {Ghirri}\ \emph {et~al.}(2016)\citenamefont {Ghirri},
  \citenamefont {Bonizzoni}, \citenamefont {Troiani}, \citenamefont {Buccheri},
  \citenamefont {Beverina}, \citenamefont {Cassinese},\ and\ \citenamefont
  {Affronte}}]{GhirriPRA16}%
  \BibitemOpen
  \bibfield  {author} {\bibinfo {author} {\bibfnamefont {A.}~\bibnamefont
  {Ghirri}}, \bibinfo {author} {\bibfnamefont {C.}~\bibnamefont {Bonizzoni}},
  \bibinfo {author} {\bibfnamefont {F.}~\bibnamefont {Troiani}}, \bibinfo
  {author} {\bibfnamefont {N.}~\bibnamefont {Buccheri}}, \bibinfo {author}
  {\bibfnamefont {L.}~\bibnamefont {Beverina}}, \bibinfo {author}
  {\bibfnamefont {A.}~\bibnamefont {Cassinese}}, \ and\ \bibinfo {author}
  {\bibfnamefont {M.}~\bibnamefont {Affronte}},\ }\href {\doibase
  10.1103/PhysRevA.93.063855} {\bibfield  {journal} {\bibinfo  {journal} {Phys.
  Rev. A}\ }\textbf {\bibinfo {volume} {93}},\ \bibinfo {pages} {063855}
  (\bibinfo {year} {2016})}\BibitemShut {NoStop}%
\bibitem [{\citenamefont {Bonizzoni}\ \emph {et~al.}(2017)\citenamefont
  {Bonizzoni}, \citenamefont {Ghirri}, \citenamefont {Atzori}, \citenamefont
  {Sorace}, \citenamefont {Sessoli},\ and\ \citenamefont
  {Affronte}}]{BonizzoniScieRep17}%
  \BibitemOpen
  \bibfield  {author} {\bibinfo {author} {\bibfnamefont {C.}~\bibnamefont
  {Bonizzoni}}, \bibinfo {author} {\bibfnamefont {A.}~\bibnamefont {Ghirri}},
  \bibinfo {author} {\bibfnamefont {M.}~\bibnamefont {Atzori}}, \bibinfo
  {author} {\bibfnamefont {L.}~\bibnamefont {Sorace}}, \bibinfo {author}
  {\bibfnamefont {R.}~\bibnamefont {Sessoli}}, \ and\ \bibinfo {author}
  {\bibfnamefont {M.}~\bibnamefont {Affronte}},\ }\href {\doibase
  10.1038/s41598-017-13271-w} {\bibfield  {journal} {\bibinfo  {journal}
  {Scientific Reports}\ }\textbf {\bibinfo {volume} {7}},\ \bibinfo {pages}
  {13096} (\bibinfo {year} {2017})}\BibitemShut {NoStop}%
\bibitem [{\citenamefont {Velluire-Pellat}\ \emph {et~al.}(2023)\citenamefont
  {Velluire-Pellat}, \citenamefont {Mar{\'e}chal}, \citenamefont {Moulonguet},
  \citenamefont {Sa{\"\i}z}, \citenamefont {M{\'e}nard}, \citenamefont
  {Kozlov}, \citenamefont {Cou{\"e}do}, \citenamefont {Amari}, \citenamefont
  {Medous}, \citenamefont {Paris}, \citenamefont {Hostein}, \citenamefont
  {Lesueur}, \citenamefont {Feuillet-Palma},\ and\ \citenamefont
  {Bergeal}}]{Velluire-PellatScieRep23}%
  \BibitemOpen
  \bibfield  {author} {\bibinfo {author} {\bibfnamefont {Z.}~\bibnamefont
  {Velluire-Pellat}}, \bibinfo {author} {\bibfnamefont {E.}~\bibnamefont
  {Mar{\'e}chal}}, \bibinfo {author} {\bibfnamefont {N.}~\bibnamefont
  {Moulonguet}}, \bibinfo {author} {\bibfnamefont {G.}~\bibnamefont
  {Sa{\"\i}z}}, \bibinfo {author} {\bibfnamefont {G.~C.}\ \bibnamefont
  {M{\'e}nard}}, \bibinfo {author} {\bibfnamefont {S.}~\bibnamefont {Kozlov}},
  \bibinfo {author} {\bibfnamefont {F.}~\bibnamefont {Cou{\"e}do}}, \bibinfo
  {author} {\bibfnamefont {P.}~\bibnamefont {Amari}}, \bibinfo {author}
  {\bibfnamefont {C.}~\bibnamefont {Medous}}, \bibinfo {author} {\bibfnamefont
  {J.}~\bibnamefont {Paris}}, \bibinfo {author} {\bibfnamefont
  {R.}~\bibnamefont {Hostein}}, \bibinfo {author} {\bibfnamefont
  {J.}~\bibnamefont {Lesueur}}, \bibinfo {author} {\bibfnamefont
  {C.}~\bibnamefont {Feuillet-Palma}}, \ and\ \bibinfo {author} {\bibfnamefont
  {N.}~\bibnamefont {Bergeal}},\ }\href
  {https://doi.org/10.1038/s41598-023-41472-z} {\bibfield  {journal} {\bibinfo
  {journal} {Scientific Reports}\ }\textbf {\bibinfo {volume} {13}},\ \bibinfo
  {pages} {14366} (\bibinfo {year} {2023})}\BibitemShut {NoStop}%
\bibitem [{\citenamefont {Gurevich}\ and\ \citenamefont
  {Melkov}(1996)}]{GuerevichMelkov}%
  \BibitemOpen
  \bibfield  {author} {\bibinfo {author} {\bibfnamefont {A.}~\bibnamefont
  {Gurevich}}\ and\ \bibinfo {author} {\bibfnamefont {G.}~\bibnamefont
  {Melkov}},\ }\href@noop {} {\emph {\bibinfo {title} {Magnetization
  Oscillations and Waves (1st ed.)}}}\ (\bibinfo  {publisher} {CRC Press.},\
  \bibinfo {year} {1996})\BibitemShut {NoStop}%
\bibitem [{\citenamefont {Pirro}\ \emph {et~al.}(2021)\citenamefont {Pirro},
  \citenamefont {Vasyuchka}, \citenamefont {Serga},\ and\ \citenamefont
  {Hillebrands}}]{PirroNatRevMater21}%
  \BibitemOpen
  \bibfield  {author} {\bibinfo {author} {\bibfnamefont {P.}~\bibnamefont
  {Pirro}}, \bibinfo {author} {\bibfnamefont {V.~I.}\ \bibnamefont
  {Vasyuchka}}, \bibinfo {author} {\bibfnamefont {A.~A.}\ \bibnamefont
  {Serga}}, \ and\ \bibinfo {author} {\bibfnamefont {B.}~\bibnamefont
  {Hillebrands}},\ }\href {https://doi.org/10.1038/s41578-021-00332-w}
  {\bibfield  {journal} {\bibinfo  {journal} {Nature Reviews Materials}\
  }\textbf {\bibinfo {volume} {6}},\ \bibinfo {pages} {1114} (\bibinfo {year}
  {2021})}\BibitemShut {NoStop}%
\bibitem [{sup()}]{supplementary}%
  \BibitemOpen
  \href@noop {} {\bibinfo  {journal} {Supplemental material}\ }\BibitemShut
  {NoStop}%
\bibitem [{\citenamefont {Kharlan}\ \emph {et~al.}(2023)\citenamefont
  {Kharlan}, \citenamefont {Sobucki}, \citenamefont {Szulc}, \citenamefont
  {Memarzadeh},\ and\ \citenamefont {Klos}}]{KharlanArxiv23}%
  \BibitemOpen
\bibfield  {journal} {  }\bibfield  {author} {\bibinfo {author} {\bibfnamefont
  {J.}~\bibnamefont {Kharlan}}, \bibinfo {author} {\bibfnamefont
  {K.}~\bibnamefont {Sobucki}}, \bibinfo {author} {\bibfnamefont
  {K.}~\bibnamefont {Szulc}}, \bibinfo {author} {\bibfnamefont
  {S.}~\bibnamefont {Memarzadeh}}, \ and\ \bibinfo {author} {\bibfnamefont
  {J.~W.}\ \bibnamefont {Klos}},\ }\href@noop {} {\enquote {\bibinfo {title}
  {Spin wave confinement in hybrid superconductor-ferrimagnet nanostructure},}\
  } (\bibinfo {year} {2023}),\ \Eprint {http://arxiv.org/abs/2312.13029}
  {arXiv:2312.13029 [cond-mat.mes-hall]} \BibitemShut {NoStop}%
\bibitem [{\citenamefont {Zhou}\ \emph {et~al.}(2024)\citenamefont {Zhou},
  \citenamefont {Ye}, \citenamefont {Bai},\ and\ \citenamefont
  {Yu}}]{ZhouArxiv24}%
  \BibitemOpen
  \bibfield  {author} {\bibinfo {author} {\bibfnamefont {X.-H.}\ \bibnamefont
  {Zhou}}, \bibinfo {author} {\bibfnamefont {X.}~\bibnamefont {Ye}}, \bibinfo
  {author} {\bibfnamefont {L.}~\bibnamefont {Bai}}, \ and\ \bibinfo {author}
  {\bibfnamefont {T.}~\bibnamefont {Yu}},\ }\href@noop {} {\enquote {\bibinfo
  {title} {Enhancement of magnon transport by superconductor meissner
  screening},}\ } (\bibinfo {year} {2024}),\ \Eprint
  {http://arxiv.org/abs/2404.02598} {arXiv:2404.02598 [cond-mat.mes-hall]}
  \BibitemShut {NoStop}%
\bibitem [{\citenamefont {Chumak}\ \emph {et~al.}(2022)\citenamefont {Chumak},
  \citenamefont {Kabos}, \citenamefont {Wu}, \citenamefont {Abert},
  \citenamefont {Adelmann}, \citenamefont {Adeyeye}, \citenamefont {Åkerman},
  \citenamefont {Aliev}, \citenamefont {Anane}, \citenamefont {Awad},
  \citenamefont {Back}, \citenamefont {Barman}, \citenamefont {Bauer},
  \citenamefont {Becherer}, \citenamefont {Beginin}, \citenamefont
  {Bittencourt}, \citenamefont {Blanter}, \citenamefont {Bortolotti},
  \citenamefont {Boventer}, \citenamefont {Bozhko}, \citenamefont {Bunyaev},
  \citenamefont {Carmiggelt}, \citenamefont {Cheenikundil}, \citenamefont
  {Ciubotaru}, \citenamefont {Cotofana}, \citenamefont {Csaba}, \citenamefont
  {Dobrovolskiy}, \citenamefont {Dubs}, \citenamefont {Elyasi}, \citenamefont
  {Fripp}, \citenamefont {Fulara}, \citenamefont {Golovchanskiy}, \citenamefont
  {Gonzalez-Ballestero}, \citenamefont {Graczyk}, \citenamefont {Grundler},
  \citenamefont {Gruszecki}, \citenamefont {Gubbiotti}, \citenamefont
  {Guslienko}, \citenamefont {Haldar}, \citenamefont {Hamdioui}, \citenamefont
  {Hertel}, \citenamefont {Hillebrands}, \citenamefont {Hioki}, \citenamefont
  {Houshang}, \citenamefont {Hu}, \citenamefont {Huebl}, \citenamefont {Huth},
  \citenamefont {Iacocca}, \citenamefont {Jungfleisch}, \citenamefont
  {Kakazei}, \citenamefont {Khitun}, \citenamefont {Khymyn}, \citenamefont
  {Kikkawa}, \citenamefont {Kläui}, \citenamefont {Klein}, \citenamefont
  {Kłos}, \citenamefont {Knauer}, \citenamefont {Koraltan}, \citenamefont
  {Kostylev}, \citenamefont {Krawczyk}, \citenamefont {Krivorotov},
  \citenamefont {Kruglyak}, \citenamefont {Lachance-Quirion}, \citenamefont
  {Ladak}, \citenamefont {Lebrun}, \citenamefont {Li}, \citenamefont {Lindner},
  \citenamefont {Macêdo}, \citenamefont {Mayr}, \citenamefont {Melkov},
  \citenamefont {Mieszczak}, \citenamefont {Nakamura}, \citenamefont {Nembach},
  \citenamefont {Nikitin}, \citenamefont {Nikitov}, \citenamefont {Novosad},
  \citenamefont {Otálora}, \citenamefont {Otani}, \citenamefont {Papp},
  \citenamefont {Pigeau}, \citenamefont {Pirro}, \citenamefont {Porod},
  \citenamefont {Porrati}, \citenamefont {Qin}, \citenamefont {Rana},
  \citenamefont {Reimann}, \citenamefont {Riente}, \citenamefont
  {Romero-Isart}, \citenamefont {Ross}, \citenamefont {Sadovnikov},
  \citenamefont {Safin}, \citenamefont {Saitoh}, \citenamefont {Schmidt},
  \citenamefont {Schultheiss}, \citenamefont {Schultheiss}, \citenamefont
  {Serga}, \citenamefont {Sharma}, \citenamefont {Shaw}, \citenamefont {Suess},
  \citenamefont {Surzhenko}, \citenamefont {Szulc}, \citenamefont {Taniguchi},
  \citenamefont {Urbánek}, \citenamefont {Usami}, \citenamefont {Ustinov},
  \citenamefont {van~der Sar}, \citenamefont {van Dijken}, \citenamefont
  {Vasyuchka}, \citenamefont {Verba}, \citenamefont {Kusminskiy}, \citenamefont
  {Wang}, \citenamefont {Weides}, \citenamefont {Weiler}, \citenamefont
  {Wintz}, \citenamefont {Wolski},\ and\ \citenamefont {Zhang}}]{ChumakIEEE22}%
  \BibitemOpen
  \bibfield  {author} {\bibinfo {author} {\bibfnamefont {A.~V.}\ \bibnamefont
  {Chumak}}, \bibinfo {author} {\bibfnamefont {P.}~\bibnamefont {Kabos}},
  \bibinfo {author} {\bibfnamefont {M.}~\bibnamefont {Wu}}, \bibinfo {author}
  {\bibfnamefont {C.}~\bibnamefont {Abert}}, \bibinfo {author} {\bibfnamefont
  {C.}~\bibnamefont {Adelmann}}, \bibinfo {author} {\bibfnamefont {A.~O.}\
  \bibnamefont {Adeyeye}}, \bibinfo {author} {\bibfnamefont {J.}~\bibnamefont
  {Åkerman}}, \bibinfo {author} {\bibfnamefont {F.~G.}\ \bibnamefont {Aliev}},
  \bibinfo {author} {\bibfnamefont {A.}~\bibnamefont {Anane}}, \bibinfo
  {author} {\bibfnamefont {A.}~\bibnamefont {Awad}}, \bibinfo {author}
  {\bibfnamefont {C.~H.}\ \bibnamefont {Back}}, \bibinfo {author}
  {\bibfnamefont {A.}~\bibnamefont {Barman}}, \bibinfo {author} {\bibfnamefont
  {G.~E.~W.}\ \bibnamefont {Bauer}}, \bibinfo {author} {\bibfnamefont
  {M.}~\bibnamefont {Becherer}}, \bibinfo {author} {\bibfnamefont {E.~N.}\
  \bibnamefont {Beginin}}, \bibinfo {author} {\bibfnamefont {V.~A. S.~V.}\
  \bibnamefont {Bittencourt}}, \bibinfo {author} {\bibfnamefont {Y.~M.}\
  \bibnamefont {Blanter}}, \bibinfo {author} {\bibfnamefont {P.}~\bibnamefont
  {Bortolotti}}, \bibinfo {author} {\bibfnamefont {I.}~\bibnamefont
  {Boventer}}, \bibinfo {author} {\bibfnamefont {D.~A.}\ \bibnamefont
  {Bozhko}}, \bibinfo {author} {\bibfnamefont {S.~A.}\ \bibnamefont {Bunyaev}},
  \bibinfo {author} {\bibfnamefont {J.~J.}\ \bibnamefont {Carmiggelt}},
  \bibinfo {author} {\bibfnamefont {R.~R.}\ \bibnamefont {Cheenikundil}},
  \bibinfo {author} {\bibfnamefont {F.}~\bibnamefont {Ciubotaru}}, \bibinfo
  {author} {\bibfnamefont {S.}~\bibnamefont {Cotofana}}, \bibinfo {author}
  {\bibfnamefont {G.}~\bibnamefont {Csaba}}, \bibinfo {author} {\bibfnamefont
  {O.~V.}\ \bibnamefont {Dobrovolskiy}}, \bibinfo {author} {\bibfnamefont
  {C.}~\bibnamefont {Dubs}}, \bibinfo {author} {\bibfnamefont {M.}~\bibnamefont
  {Elyasi}}, \bibinfo {author} {\bibfnamefont {K.~G.}\ \bibnamefont {Fripp}},
  \bibinfo {author} {\bibfnamefont {H.}~\bibnamefont {Fulara}}, \bibinfo
  {author} {\bibfnamefont {I.~A.}\ \bibnamefont {Golovchanskiy}}, \bibinfo
  {author} {\bibfnamefont {C.}~\bibnamefont {Gonzalez-Ballestero}}, \bibinfo
  {author} {\bibfnamefont {P.}~\bibnamefont {Graczyk}}, \bibinfo {author}
  {\bibfnamefont {D.}~\bibnamefont {Grundler}}, \bibinfo {author}
  {\bibfnamefont {P.}~\bibnamefont {Gruszecki}}, \bibinfo {author}
  {\bibfnamefont {G.}~\bibnamefont {Gubbiotti}}, \bibinfo {author}
  {\bibfnamefont {K.}~\bibnamefont {Guslienko}}, \bibinfo {author}
  {\bibfnamefont {A.}~\bibnamefont {Haldar}}, \bibinfo {author} {\bibfnamefont
  {S.}~\bibnamefont {Hamdioui}}, \bibinfo {author} {\bibfnamefont
  {R.}~\bibnamefont {Hertel}}, \bibinfo {author} {\bibfnamefont
  {B.}~\bibnamefont {Hillebrands}}, \bibinfo {author} {\bibfnamefont
  {T.}~\bibnamefont {Hioki}}, \bibinfo {author} {\bibfnamefont
  {A.}~\bibnamefont {Houshang}}, \bibinfo {author} {\bibfnamefont {C.-M.}\
  \bibnamefont {Hu}}, \bibinfo {author} {\bibfnamefont {H.}~\bibnamefont
  {Huebl}}, \bibinfo {author} {\bibfnamefont {M.}~\bibnamefont {Huth}},
  \bibinfo {author} {\bibfnamefont {E.}~\bibnamefont {Iacocca}}, \bibinfo
  {author} {\bibfnamefont {M.~B.}\ \bibnamefont {Jungfleisch}}, \bibinfo
  {author} {\bibfnamefont {G.~N.}\ \bibnamefont {Kakazei}}, \bibinfo {author}
  {\bibfnamefont {A.}~\bibnamefont {Khitun}}, \bibinfo {author} {\bibfnamefont
  {R.}~\bibnamefont {Khymyn}}, \bibinfo {author} {\bibfnamefont
  {T.}~\bibnamefont {Kikkawa}}, \bibinfo {author} {\bibfnamefont
  {M.}~\bibnamefont {Kläui}}, \bibinfo {author} {\bibfnamefont
  {O.}~\bibnamefont {Klein}}, \bibinfo {author} {\bibfnamefont {J.~W.}\
  \bibnamefont {Kłos}}, \bibinfo {author} {\bibfnamefont {S.}~\bibnamefont
  {Knauer}}, \bibinfo {author} {\bibfnamefont {S.}~\bibnamefont {Koraltan}},
  \bibinfo {author} {\bibfnamefont {M.}~\bibnamefont {Kostylev}}, \bibinfo
  {author} {\bibfnamefont {M.}~\bibnamefont {Krawczyk}}, \bibinfo {author}
  {\bibfnamefont {I.~N.}\ \bibnamefont {Krivorotov}}, \bibinfo {author}
  {\bibfnamefont {V.~V.}\ \bibnamefont {Kruglyak}}, \bibinfo {author}
  {\bibfnamefont {D.}~\bibnamefont {Lachance-Quirion}}, \bibinfo {author}
  {\bibfnamefont {S.}~\bibnamefont {Ladak}}, \bibinfo {author} {\bibfnamefont
  {R.}~\bibnamefont {Lebrun}}, \bibinfo {author} {\bibfnamefont
  {Y.}~\bibnamefont {Li}}, \bibinfo {author} {\bibfnamefont {M.}~\bibnamefont
  {Lindner}}, \bibinfo {author} {\bibfnamefont {R.}~\bibnamefont {Macêdo}},
  \bibinfo {author} {\bibfnamefont {S.}~\bibnamefont {Mayr}}, \bibinfo {author}
  {\bibfnamefont {G.~A.}\ \bibnamefont {Melkov}}, \bibinfo {author}
  {\bibfnamefont {S.}~\bibnamefont {Mieszczak}}, \bibinfo {author}
  {\bibfnamefont {Y.}~\bibnamefont {Nakamura}}, \bibinfo {author}
  {\bibfnamefont {H.~T.}\ \bibnamefont {Nembach}}, \bibinfo {author}
  {\bibfnamefont {A.~A.}\ \bibnamefont {Nikitin}}, \bibinfo {author}
  {\bibfnamefont {S.~A.}\ \bibnamefont {Nikitov}}, \bibinfo {author}
  {\bibfnamefont {V.}~\bibnamefont {Novosad}}, \bibinfo {author} {\bibfnamefont
  {J.~A.}\ \bibnamefont {Otálora}}, \bibinfo {author} {\bibfnamefont
  {Y.}~\bibnamefont {Otani}}, \bibinfo {author} {\bibfnamefont
  {A.}~\bibnamefont {Papp}}, \bibinfo {author} {\bibfnamefont {B.}~\bibnamefont
  {Pigeau}}, \bibinfo {author} {\bibfnamefont {P.}~\bibnamefont {Pirro}},
  \bibinfo {author} {\bibfnamefont {W.}~\bibnamefont {Porod}}, \bibinfo
  {author} {\bibfnamefont {F.}~\bibnamefont {Porrati}}, \bibinfo {author}
  {\bibfnamefont {H.}~\bibnamefont {Qin}}, \bibinfo {author} {\bibfnamefont
  {B.}~\bibnamefont {Rana}}, \bibinfo {author} {\bibfnamefont {T.}~\bibnamefont
  {Reimann}}, \bibinfo {author} {\bibfnamefont {F.}~\bibnamefont {Riente}},
  \bibinfo {author} {\bibfnamefont {O.}~\bibnamefont {Romero-Isart}}, \bibinfo
  {author} {\bibfnamefont {A.}~\bibnamefont {Ross}}, \bibinfo {author}
  {\bibfnamefont {A.~V.}\ \bibnamefont {Sadovnikov}}, \bibinfo {author}
  {\bibfnamefont {A.~R.}\ \bibnamefont {Safin}}, \bibinfo {author}
  {\bibfnamefont {E.}~\bibnamefont {Saitoh}}, \bibinfo {author} {\bibfnamefont
  {G.}~\bibnamefont {Schmidt}}, \bibinfo {author} {\bibfnamefont
  {H.}~\bibnamefont {Schultheiss}}, \bibinfo {author} {\bibfnamefont
  {K.}~\bibnamefont {Schultheiss}}, \bibinfo {author} {\bibfnamefont {A.~A.}\
  \bibnamefont {Serga}}, \bibinfo {author} {\bibfnamefont {S.}~\bibnamefont
  {Sharma}}, \bibinfo {author} {\bibfnamefont {J.~M.}\ \bibnamefont {Shaw}},
  \bibinfo {author} {\bibfnamefont {D.}~\bibnamefont {Suess}}, \bibinfo
  {author} {\bibfnamefont {O.}~\bibnamefont {Surzhenko}}, \bibinfo {author}
  {\bibfnamefont {K.}~\bibnamefont {Szulc}}, \bibinfo {author} {\bibfnamefont
  {T.}~\bibnamefont {Taniguchi}}, \bibinfo {author} {\bibfnamefont
  {M.}~\bibnamefont {Urbánek}}, \bibinfo {author} {\bibfnamefont
  {K.}~\bibnamefont {Usami}}, \bibinfo {author} {\bibfnamefont {A.~B.}\
  \bibnamefont {Ustinov}}, \bibinfo {author} {\bibfnamefont {T.}~\bibnamefont
  {van~der Sar}}, \bibinfo {author} {\bibfnamefont {S.}~\bibnamefont {van
  Dijken}}, \bibinfo {author} {\bibfnamefont {V.~I.}\ \bibnamefont
  {Vasyuchka}}, \bibinfo {author} {\bibfnamefont {R.}~\bibnamefont {Verba}},
  \bibinfo {author} {\bibfnamefont {S.~V.}\ \bibnamefont {Kusminskiy}},
  \bibinfo {author} {\bibfnamefont {Q.}~\bibnamefont {Wang}}, \bibinfo {author}
  {\bibfnamefont {M.}~\bibnamefont {Weides}}, \bibinfo {author} {\bibfnamefont
  {M.}~\bibnamefont {Weiler}}, \bibinfo {author} {\bibfnamefont
  {S.}~\bibnamefont {Wintz}}, \bibinfo {author} {\bibfnamefont {S.~P.}\
  \bibnamefont {Wolski}}, \ and\ \bibinfo {author} {\bibfnamefont
  {X.}~\bibnamefont {Zhang}},\ }\href {\doibase 10.1109/TMAG.2022.3149664}
  {\bibfield  {journal} {\bibinfo  {journal} {IEEE Transactions on Magnetics}\
  }\textbf {\bibinfo {volume} {58}},\ \bibinfo {pages} {1} (\bibinfo {year}
  {2022})}\BibitemShut {NoStop}%
\bibitem [{\citenamefont {Lachance-Quirion}\ \emph {et~al.}(2019)\citenamefont
  {Lachance-Quirion}, \citenamefont {Tabuchi}, \citenamefont {Gloppe},
  \citenamefont {Usami},\ and\ \citenamefont
  {Nakamura}}]{Lachance-QuirionApplPhysExp19}%
  \BibitemOpen
  \bibfield  {author} {\bibinfo {author} {\bibfnamefont {D.}~\bibnamefont
  {Lachance-Quirion}}, \bibinfo {author} {\bibfnamefont {Y.}~\bibnamefont
  {Tabuchi}}, \bibinfo {author} {\bibfnamefont {A.}~\bibnamefont {Gloppe}},
  \bibinfo {author} {\bibfnamefont {K.}~\bibnamefont {Usami}}, \ and\ \bibinfo
  {author} {\bibfnamefont {Y.}~\bibnamefont {Nakamura}},\ }\href@noop {}
  {\bibfield  {journal} {\bibinfo  {journal} {Appl. Phys. Express}\ }\textbf
  {\bibinfo {volume} {12}} (\bibinfo {year} {2019})}\BibitemShut {NoStop}%
\bibitem [{\citenamefont {{Zare Rameshti}}\ \emph {et~al.}(2022)\citenamefont
  {{Zare Rameshti}}, \citenamefont {{Viola Kusminskiy}}, \citenamefont {Haigh},
  \citenamefont {Usami}, \citenamefont {Lachance-Quirion}, \citenamefont
  {Nakamura}, \citenamefont {Hu}, \citenamefont {Tang}, \citenamefont {Bauer},\
  and\ \citenamefont {Blanter}}]{RameshtiPhysRep22}%
  \BibitemOpen
  \bibfield  {author} {\bibinfo {author} {\bibfnamefont {B.}~\bibnamefont
  {{Zare Rameshti}}}, \bibinfo {author} {\bibfnamefont {S.}~\bibnamefont
  {{Viola Kusminskiy}}}, \bibinfo {author} {\bibfnamefont {J.~A.}\ \bibnamefont
  {Haigh}}, \bibinfo {author} {\bibfnamefont {K.}~\bibnamefont {Usami}},
  \bibinfo {author} {\bibfnamefont {D.}~\bibnamefont {Lachance-Quirion}},
  \bibinfo {author} {\bibfnamefont {Y.}~\bibnamefont {Nakamura}}, \bibinfo
  {author} {\bibfnamefont {C.-M.}\ \bibnamefont {Hu}}, \bibinfo {author}
  {\bibfnamefont {H.~X.}\ \bibnamefont {Tang}}, \bibinfo {author}
  {\bibfnamefont {G.~E.}\ \bibnamefont {Bauer}}, \ and\ \bibinfo {author}
  {\bibfnamefont {Y.~M.}\ \bibnamefont {Blanter}},\ }\href
  {https://www.sciencedirect.com/science/article/pii/S0370157322002460}
  {\bibfield  {journal} {\bibinfo  {journal} {Physics Reports}\ }\textbf
  {\bibinfo {volume} {979}},\ \bibinfo {pages} {1} (\bibinfo {year}
  {2022})}\BibitemShut {NoStop}%
\bibitem [{\citenamefont {Huebl}\ \emph {et~al.}(2013)\citenamefont {Huebl},
  \citenamefont {Zollitsch}, \citenamefont {Lotze}, \citenamefont {Hocke},
  \citenamefont {Greifenstein}, \citenamefont {Marx}, \citenamefont {Gross},\
  and\ \citenamefont {Goennenwein}}]{HueblPRL13}%
  \BibitemOpen
  \bibfield  {author} {\bibinfo {author} {\bibfnamefont {H.}~\bibnamefont
  {Huebl}}, \bibinfo {author} {\bibfnamefont {C.~W.}\ \bibnamefont
  {Zollitsch}}, \bibinfo {author} {\bibfnamefont {J.}~\bibnamefont {Lotze}},
  \bibinfo {author} {\bibfnamefont {F.}~\bibnamefont {Hocke}}, \bibinfo
  {author} {\bibfnamefont {M.}~\bibnamefont {Greifenstein}}, \bibinfo {author}
  {\bibfnamefont {A.}~\bibnamefont {Marx}}, \bibinfo {author} {\bibfnamefont
  {R.}~\bibnamefont {Gross}}, \ and\ \bibinfo {author} {\bibfnamefont
  {S.~T.~B.}\ \bibnamefont {Goennenwein}},\ }\href {\doibase
  10.1103/PhysRevLett.111.127003} {\bibfield  {journal} {\bibinfo  {journal}
  {Phys. Rev. Lett.}\ }\textbf {\bibinfo {volume} {111}},\ \bibinfo {pages}
  {127003} (\bibinfo {year} {2013})}\BibitemShut {NoStop}%
\bibitem [{\citenamefont {Morris}\ \emph {et~al.}(2017)\citenamefont {Morris},
  \citenamefont {van Loo}, \citenamefont {Kosen},\ and\ \citenamefont
  {Karenowska}}]{MorrisScieRep17}%
  \BibitemOpen
  \bibfield  {author} {\bibinfo {author} {\bibfnamefont {R.~G.~E.}\
  \bibnamefont {Morris}}, \bibinfo {author} {\bibfnamefont {A.~F.}\
  \bibnamefont {van Loo}}, \bibinfo {author} {\bibfnamefont {S.}~\bibnamefont
  {Kosen}}, \ and\ \bibinfo {author} {\bibfnamefont {A.~D.}\ \bibnamefont
  {Karenowska}},\ }\href {\doibase 10.1038/s41598-017-11835-4} {\bibfield
  {journal} {\bibinfo  {journal} {Scientific Reports}\ }\textbf {\bibinfo
  {volume} {7}},\ \bibinfo {pages} {11511} (\bibinfo {year}
  {2017})}\BibitemShut {NoStop}%
\bibitem [{\citenamefont {Hou}\ and\ \citenamefont {Liu}(2019)}]{HouPRL19}%
  \BibitemOpen
  \bibfield  {author} {\bibinfo {author} {\bibfnamefont {J.~T.}\ \bibnamefont
  {Hou}}\ and\ \bibinfo {author} {\bibfnamefont {L.}~\bibnamefont {Liu}},\
  }\href {https://link.aps.org/doi/10.1103/PhysRevLett.123.107702} {\bibfield
  {journal} {\bibinfo  {journal} {Phys. Rev. Lett.}\ }\textbf {\bibinfo
  {volume} {123}},\ \bibinfo {pages} {107702} (\bibinfo {year}
  {2019})}\BibitemShut {NoStop}%
\bibitem [{\citenamefont {Li}\ \emph {et~al.}(2019)\citenamefont {Li},
  \citenamefont {Polakovic}, \citenamefont {Wang}, \citenamefont {Xu},
  \citenamefont {Lendinez}, \citenamefont {Zhang}, \citenamefont {Ding},
  \citenamefont {Khaire}, \citenamefont {Saglam}, \citenamefont {Divan},
  \citenamefont {Pearson}, \citenamefont {Kwok}, \citenamefont {Xiao},
  \citenamefont {Novosad}, \citenamefont {Hoffmann},\ and\ \citenamefont
  {Zhang}}]{LiPRL19}%
  \BibitemOpen
  \bibfield  {author} {\bibinfo {author} {\bibfnamefont {Y.}~\bibnamefont
  {Li}}, \bibinfo {author} {\bibfnamefont {T.}~\bibnamefont {Polakovic}},
  \bibinfo {author} {\bibfnamefont {Y.-L.}\ \bibnamefont {Wang}}, \bibinfo
  {author} {\bibfnamefont {J.}~\bibnamefont {Xu}}, \bibinfo {author}
  {\bibfnamefont {S.}~\bibnamefont {Lendinez}}, \bibinfo {author}
  {\bibfnamefont {Z.}~\bibnamefont {Zhang}}, \bibinfo {author} {\bibfnamefont
  {J.}~\bibnamefont {Ding}}, \bibinfo {author} {\bibfnamefont {T.}~\bibnamefont
  {Khaire}}, \bibinfo {author} {\bibfnamefont {H.}~\bibnamefont {Saglam}},
  \bibinfo {author} {\bibfnamefont {R.}~\bibnamefont {Divan}}, \bibinfo
  {author} {\bibfnamefont {J.}~\bibnamefont {Pearson}}, \bibinfo {author}
  {\bibfnamefont {W.-K.}\ \bibnamefont {Kwok}}, \bibinfo {author}
  {\bibfnamefont {Z.}~\bibnamefont {Xiao}}, \bibinfo {author} {\bibfnamefont
  {V.}~\bibnamefont {Novosad}}, \bibinfo {author} {\bibfnamefont
  {A.}~\bibnamefont {Hoffmann}}, \ and\ \bibinfo {author} {\bibfnamefont
  {W.}~\bibnamefont {Zhang}},\ }\href {\doibase 10.1103/PhysRevLett.123.107701}
  {\bibfield  {journal} {\bibinfo  {journal} {Phys. Rev. Lett.}\ }\textbf
  {\bibinfo {volume} {123}},\ \bibinfo {pages} {107701} (\bibinfo {year}
  {2019})}\BibitemShut {NoStop}%
\bibitem [{\citenamefont {Mandal}\ \emph {et~al.}(2020)\citenamefont {Mandal},
  \citenamefont {Kapoor}, \citenamefont {Ghosh}, \citenamefont {Jesudasan},
  \citenamefont {Manni}, \citenamefont {Thamizhavel}, \citenamefont
  {Raychaudhuri}, \citenamefont {Singh},\ and\ \citenamefont
  {Deshmukh}}]{MandalAPL20}%
  \BibitemOpen
  \bibfield  {author} {\bibinfo {author} {\bibfnamefont {S.}~\bibnamefont
  {Mandal}}, \bibinfo {author} {\bibfnamefont {L.~N.}\ \bibnamefont {Kapoor}},
  \bibinfo {author} {\bibfnamefont {S.}~\bibnamefont {Ghosh}}, \bibinfo
  {author} {\bibfnamefont {J.}~\bibnamefont {Jesudasan}}, \bibinfo {author}
  {\bibfnamefont {S.}~\bibnamefont {Manni}}, \bibinfo {author} {\bibfnamefont
  {A.}~\bibnamefont {Thamizhavel}}, \bibinfo {author} {\bibfnamefont
  {P.}~\bibnamefont {Raychaudhuri}}, \bibinfo {author} {\bibfnamefont
  {V.}~\bibnamefont {Singh}}, \ and\ \bibinfo {author} {\bibfnamefont {M.~M.}\
  \bibnamefont {Deshmukh}},\ }\href {https://doi.org/10.1063/5.0029112}
  {\bibfield  {journal} {\bibinfo  {journal} {Applied Physics Letters}\
  }\textbf {\bibinfo {volume} {117}},\ \bibinfo {pages} {263101} (\bibinfo
  {year} {2020})}\BibitemShut {NoStop}%
\bibitem [{\citenamefont {Haygood}\ \emph {et~al.}(2021)\citenamefont
  {Haygood}, \citenamefont {Pufall}, \citenamefont {Edwards}, \citenamefont
  {Shaw},\ and\ \citenamefont {Rippard}}]{HaygoodPRAppl21}%
  \BibitemOpen
  \bibfield  {author} {\bibinfo {author} {\bibfnamefont {I.}~\bibnamefont
  {Haygood}}, \bibinfo {author} {\bibfnamefont {M.}~\bibnamefont {Pufall}},
  \bibinfo {author} {\bibfnamefont {E.}~\bibnamefont {Edwards}}, \bibinfo
  {author} {\bibfnamefont {J.~M.}\ \bibnamefont {Shaw}}, \ and\ \bibinfo
  {author} {\bibfnamefont {W.}~\bibnamefont {Rippard}},\ }\href {\doibase
  10.1103/PhysRevApplied.15.054021} {\bibfield  {journal} {\bibinfo  {journal}
  {Phys. Rev. Appl.}\ }\textbf {\bibinfo {volume} {15}},\ \bibinfo {pages}
  {054021} (\bibinfo {year} {2021})}\BibitemShut {NoStop}%
\bibitem [{\citenamefont {Baity}\ \emph {et~al.}(2021)\citenamefont {Baity},
  \citenamefont {Bozhko}, \citenamefont {Mac{\^e}do}, \citenamefont {Smith},
  \citenamefont {Holland}, \citenamefont {Danilin}, \citenamefont {Seferai},
  \citenamefont {Barbosa}, \citenamefont {Peroor}, \citenamefont {Goldman},
  \citenamefont {Nasti}, \citenamefont {Paul}, \citenamefont {Hadfield},
  \citenamefont {McVitie},\ and\ \citenamefont {Weides}}]{BaityAPL21}%
  \BibitemOpen
  \bibfield  {author} {\bibinfo {author} {\bibfnamefont {P.~G.}\ \bibnamefont
  {Baity}}, \bibinfo {author} {\bibfnamefont {D.~A.}\ \bibnamefont {Bozhko}},
  \bibinfo {author} {\bibfnamefont {R.}~\bibnamefont {Mac{\^e}do}}, \bibinfo
  {author} {\bibfnamefont {W.}~\bibnamefont {Smith}}, \bibinfo {author}
  {\bibfnamefont {R.~C.}\ \bibnamefont {Holland}}, \bibinfo {author}
  {\bibfnamefont {S.}~\bibnamefont {Danilin}}, \bibinfo {author} {\bibfnamefont
  {V.}~\bibnamefont {Seferai}}, \bibinfo {author} {\bibfnamefont
  {J.}~\bibnamefont {Barbosa}}, \bibinfo {author} {\bibfnamefont {R.~R.}\
  \bibnamefont {Peroor}}, \bibinfo {author} {\bibfnamefont {S.}~\bibnamefont
  {Goldman}}, \bibinfo {author} {\bibfnamefont {U.}~\bibnamefont {Nasti}},
  \bibinfo {author} {\bibfnamefont {J.}~\bibnamefont {Paul}}, \bibinfo {author}
  {\bibfnamefont {R.~H.}\ \bibnamefont {Hadfield}}, \bibinfo {author}
  {\bibfnamefont {S.}~\bibnamefont {McVitie}}, \ and\ \bibinfo {author}
  {\bibfnamefont {M.}~\bibnamefont {Weides}},\ }\href {\doibase
  10.1063/5.0054837} {\bibfield  {journal} {\bibinfo  {journal} {Applied
  Physics Letters}\ }\textbf {\bibinfo {volume} {119}},\ \bibinfo {pages}
  {033502} (\bibinfo {year} {2021})}\BibitemShut {NoStop}%
\bibitem [{\citenamefont {Li}\ \emph {et~al.}(2022)\citenamefont {Li},
  \citenamefont {Yefremenko}, \citenamefont {Lisovenko}, \citenamefont
  {Trevillian}, \citenamefont {Polakovic}, \citenamefont {Cecil}, \citenamefont
  {Barry}, \citenamefont {Pearson}, \citenamefont {Divan}, \citenamefont
  {Tyberkevych}, \citenamefont {Chang}, \citenamefont {Welp}, \citenamefont
  {Kwok},\ and\ \citenamefont {Novosad}}]{LiPRL22}%
  \BibitemOpen
  \bibfield  {author} {\bibinfo {author} {\bibfnamefont {Y.}~\bibnamefont
  {Li}}, \bibinfo {author} {\bibfnamefont {V.~G.}\ \bibnamefont {Yefremenko}},
  \bibinfo {author} {\bibfnamefont {M.}~\bibnamefont {Lisovenko}}, \bibinfo
  {author} {\bibfnamefont {C.}~\bibnamefont {Trevillian}}, \bibinfo {author}
  {\bibfnamefont {T.}~\bibnamefont {Polakovic}}, \bibinfo {author}
  {\bibfnamefont {T.~W.}\ \bibnamefont {Cecil}}, \bibinfo {author}
  {\bibfnamefont {P.~S.}\ \bibnamefont {Barry}}, \bibinfo {author}
  {\bibfnamefont {J.}~\bibnamefont {Pearson}}, \bibinfo {author} {\bibfnamefont
  {R.}~\bibnamefont {Divan}}, \bibinfo {author} {\bibfnamefont
  {V.}~\bibnamefont {Tyberkevych}}, \bibinfo {author} {\bibfnamefont {C.~L.}\
  \bibnamefont {Chang}}, \bibinfo {author} {\bibfnamefont {U.}~\bibnamefont
  {Welp}}, \bibinfo {author} {\bibfnamefont {W.-K.}\ \bibnamefont {Kwok}}, \
  and\ \bibinfo {author} {\bibfnamefont {V.}~\bibnamefont {Novosad}},\ }\href
  {\doibase 10.1103/PhysRevLett.128.047701} {\bibfield  {journal} {\bibinfo
  {journal} {Phys. Rev. Lett.}\ }\textbf {\bibinfo {volume} {128}},\ \bibinfo
  {pages} {047701} (\bibinfo {year} {2022})}\BibitemShut {NoStop}%
\bibitem [{\citenamefont {Bøttcher}\ \emph {et~al.}(2023)\citenamefont
  {Bøttcher}, \citenamefont {Poniatowski}, \citenamefont {Grankin},
  \citenamefont {Wesson}, \citenamefont {Yan}, \citenamefont {Vool},
  \citenamefont {Galitski},\ and\ \citenamefont {Yacoby}}]{BottcherArxiv23}%
  \BibitemOpen
  \bibfield  {author} {\bibinfo {author} {\bibfnamefont {C.~G.~L.}\
  \bibnamefont {Bøttcher}}, \bibinfo {author} {\bibfnamefont {N.~R.}\
  \bibnamefont {Poniatowski}}, \bibinfo {author} {\bibfnamefont
  {A.}~\bibnamefont {Grankin}}, \bibinfo {author} {\bibfnamefont {M.~E.}\
  \bibnamefont {Wesson}}, \bibinfo {author} {\bibfnamefont {Z.}~\bibnamefont
  {Yan}}, \bibinfo {author} {\bibfnamefont {U.}~\bibnamefont {Vool}}, \bibinfo
  {author} {\bibfnamefont {V.~M.}\ \bibnamefont {Galitski}}, \ and\ \bibinfo
  {author} {\bibfnamefont {A.}~\bibnamefont {Yacoby}},\ }\href@noop {}
  {\enquote {\bibinfo {title} {Circuit qed detection of induced two-fold
  anisotropic pairing in a hybrid superconductor-ferromagnet bilayer},}\ }
  (\bibinfo {year} {2023}),\ \Eprint {http://arxiv.org/abs/2306.08043}
  {arXiv:2306.08043 [cond-mat.supr-con]} \BibitemShut {NoStop}%
\bibitem [{\citenamefont {Golovchanskiy}\ \emph
  {et~al.}(2021{\natexlab{a}})\citenamefont {Golovchanskiy}, \citenamefont
  {Abramov}, \citenamefont {Stolyarov}, \citenamefont {Weides}, \citenamefont
  {Ryazanov}, \citenamefont {Golubov}, \citenamefont {Ustinov},\ and\
  \citenamefont {Kupriyanov}}]{GolovchanskiyScAdv21}%
  \BibitemOpen
  \bibfield  {author} {\bibinfo {author} {\bibfnamefont {I.~A.}\ \bibnamefont
  {Golovchanskiy}}, \bibinfo {author} {\bibfnamefont {N.~N.}\ \bibnamefont
  {Abramov}}, \bibinfo {author} {\bibfnamefont {V.~S.}\ \bibnamefont
  {Stolyarov}}, \bibinfo {author} {\bibfnamefont {M.}~\bibnamefont {Weides}},
  \bibinfo {author} {\bibfnamefont {V.~V.}\ \bibnamefont {Ryazanov}}, \bibinfo
  {author} {\bibfnamefont {A.~A.}\ \bibnamefont {Golubov}}, \bibinfo {author}
  {\bibfnamefont {A.~V.}\ \bibnamefont {Ustinov}}, \ and\ \bibinfo {author}
  {\bibfnamefont {M.~Y.}\ \bibnamefont {Kupriyanov}},\ }\href {\doibase
  10.1126/sciadv.abe8638} {\bibfield  {journal} {\bibinfo  {journal} {Science
  Advances}\ }\textbf {\bibinfo {volume} {7}},\ \bibinfo {pages} {eabe8638}
  (\bibinfo {year} {2021}{\natexlab{a}})}\BibitemShut {NoStop}%
\bibitem [{\citenamefont {Golovchanskiy}\ \emph
  {et~al.}(2021{\natexlab{b}})\citenamefont {Golovchanskiy}, \citenamefont
  {Abramov}, \citenamefont {Stolyarov}, \citenamefont {Golubov}, \citenamefont
  {Kupriyanov}, \citenamefont {Ryazanov},\ and\ \citenamefont
  {Ustinov}}]{GolovchanskiyPRAppl21}%
  \BibitemOpen
  \bibfield  {author} {\bibinfo {author} {\bibfnamefont {I.}~\bibnamefont
  {Golovchanskiy}}, \bibinfo {author} {\bibfnamefont {N.}~\bibnamefont
  {Abramov}}, \bibinfo {author} {\bibfnamefont {V.}~\bibnamefont {Stolyarov}},
  \bibinfo {author} {\bibfnamefont {A.}~\bibnamefont {Golubov}}, \bibinfo
  {author} {\bibfnamefont {M.~Y.}\ \bibnamefont {Kupriyanov}}, \bibinfo
  {author} {\bibfnamefont {V.}~\bibnamefont {Ryazanov}}, \ and\ \bibinfo
  {author} {\bibfnamefont {A.}~\bibnamefont {Ustinov}},\ }\href {\doibase
  10.1103/PhysRevApplied.16.034029} {\bibfield  {journal} {\bibinfo  {journal}
  {Phys. Rev. Applied}\ }\textbf {\bibinfo {volume} {16}},\ \bibinfo {pages}
  {034029} (\bibinfo {year} {2021}{\natexlab{b}})}\BibitemShut {NoStop}%
\bibitem [{\citenamefont {Silaev}(2023)}]{SilaevPRB23}%
  \BibitemOpen
  \bibfield  {author} {\bibinfo {author} {\bibfnamefont {M.}~\bibnamefont
  {Silaev}},\ }\href {\doibase 10.1103/PhysRevB.107.L180503} {\bibfield
  {journal} {\bibinfo  {journal} {Phys. Rev. B}\ }\textbf {\bibinfo {volume}
  {107}},\ \bibinfo {pages} {L180503} (\bibinfo {year} {2023})}\BibitemShut
  {NoStop}%
\bibitem [{\citenamefont {Ghirri}\ \emph {et~al.}(2023)\citenamefont {Ghirri},
  \citenamefont {Bonizzoni}, \citenamefont {Maksutoglu}, \citenamefont
  {Mercurio}, \citenamefont {Di~Stefano}, \citenamefont {Savasta},\ and\
  \citenamefont {Affronte}}]{GhirriPrAppl23}%
  \BibitemOpen
  \bibfield  {author} {\bibinfo {author} {\bibfnamefont {A.}~\bibnamefont
  {Ghirri}}, \bibinfo {author} {\bibfnamefont {C.}~\bibnamefont {Bonizzoni}},
  \bibinfo {author} {\bibfnamefont {M.}~\bibnamefont {Maksutoglu}}, \bibinfo
  {author} {\bibfnamefont {A.}~\bibnamefont {Mercurio}}, \bibinfo {author}
  {\bibfnamefont {O.}~\bibnamefont {Di~Stefano}}, \bibinfo {author}
  {\bibfnamefont {S.}~\bibnamefont {Savasta}}, \ and\ \bibinfo {author}
  {\bibfnamefont {M.}~\bibnamefont {Affronte}},\ }\href {\doibase
  10.1103/PhysRevApplied.20.024039} {\bibfield  {journal} {\bibinfo  {journal}
  {Phys. Rev. Appl.}\ }\textbf {\bibinfo {volume} {20}},\ \bibinfo {pages}
  {024039} (\bibinfo {year} {2023})}\BibitemShut {NoStop}%
\bibitem [{\citenamefont {Frisk~Kockum}\ \emph {et~al.}(2019)\citenamefont
  {Frisk~Kockum}, \citenamefont {Miranowicz}, \citenamefont {De~Liberato},
  \citenamefont {Savasta},\ and\ \citenamefont {Nori}}]{Kockum2019}%
  \BibitemOpen
  \bibfield  {author} {\bibinfo {author} {\bibfnamefont {A.}~\bibnamefont
  {Frisk~Kockum}}, \bibinfo {author} {\bibfnamefont {A.}~\bibnamefont
  {Miranowicz}}, \bibinfo {author} {\bibfnamefont {S.}~\bibnamefont
  {De~Liberato}}, \bibinfo {author} {\bibfnamefont {S.}~\bibnamefont
  {Savasta}}, \ and\ \bibinfo {author} {\bibfnamefont {F.}~\bibnamefont
  {Nori}},\ }\href {\doibase 10.1038/s42254-018-0006-2} {\bibfield  {journal}
  {\bibinfo  {journal} {Nature Reviews Physics}\ }\textbf {\bibinfo {volume}
  {1}},\ \bibinfo {pages} {19} (\bibinfo {year} {2019})}\BibitemShut {NoStop}%
\bibitem [{\citenamefont {Vendik}\ \emph {et~al.}(1998)\citenamefont {Vendik},
  \citenamefont {Vendik},\ and\ \citenamefont {Kaparkov}}]{VendikIEEE98}%
  \BibitemOpen
  \bibfield  {author} {\bibinfo {author} {\bibfnamefont {O.}~\bibnamefont
  {Vendik}}, \bibinfo {author} {\bibfnamefont {I.}~\bibnamefont {Vendik}}, \
  and\ \bibinfo {author} {\bibfnamefont {D.}~\bibnamefont {Kaparkov}},\ }\href
  {\doibase 10.1109/22.668643} {\bibfield  {journal} {\bibinfo  {journal} {IEEE
  Transactions on Microwave Theory and Techniques}\ }\textbf {\bibinfo {volume}
  {46}},\ \bibinfo {pages} {469} (\bibinfo {year} {1998})}\BibitemShut
  {NoStop}%
\bibitem [{\citenamefont {Ghigo}\ \emph {et~al.}(2004)\citenamefont {Ghigo},
  \citenamefont {Botta}, \citenamefont {Chiodoni}, \citenamefont {Gerbaldo},
  \citenamefont {Gozzelino}, \citenamefont {Laviano}, \citenamefont {Minetti},
  \citenamefont {Mezzetti},\ and\ \citenamefont
  {Andreone}}]{GhigoSuperSciTech04}%
  \BibitemOpen
  \bibfield  {author} {\bibinfo {author} {\bibfnamefont {G.}~\bibnamefont
  {Ghigo}}, \bibinfo {author} {\bibfnamefont {D.}~\bibnamefont {Botta}},
  \bibinfo {author} {\bibfnamefont {A.}~\bibnamefont {Chiodoni}}, \bibinfo
  {author} {\bibfnamefont {R.}~\bibnamefont {Gerbaldo}}, \bibinfo {author}
  {\bibfnamefont {L.}~\bibnamefont {Gozzelino}}, \bibinfo {author}
  {\bibfnamefont {F.}~\bibnamefont {Laviano}}, \bibinfo {author} {\bibfnamefont
  {B.}~\bibnamefont {Minetti}}, \bibinfo {author} {\bibfnamefont
  {E.}~\bibnamefont {Mezzetti}}, \ and\ \bibinfo {author} {\bibfnamefont
  {D.}~\bibnamefont {Andreone}},\ }\href {\doibase 10.1088/0953-2048/17/8/004}
  {\bibfield  {journal} {\bibinfo  {journal} {Superconductor Science and
  Technology}\ }\textbf {\bibinfo {volume} {17}},\ \bibinfo {pages} {977}
  (\bibinfo {year} {2004})}\BibitemShut {NoStop}%
\bibitem [{\citenamefont {Prozorov}\ and\ \citenamefont
  {Giannetta}(2006)}]{ProzorovSuperScieTech06}%
  \BibitemOpen
  \bibfield  {author} {\bibinfo {author} {\bibfnamefont {R.}~\bibnamefont
  {Prozorov}}\ and\ \bibinfo {author} {\bibfnamefont {R.~W.}\ \bibnamefont
  {Giannetta}},\ }\href {https://dx.doi.org/10.1088/0953-2048/19/8/R01}
  {\bibfield  {journal} {\bibinfo  {journal} {Superconductor Science and
  Technology}\ }\textbf {\bibinfo {volume} {19}},\ \bibinfo {pages} {R41}
  (\bibinfo {year} {2006})}\BibitemShut {NoStop}%
\bibitem [{\citenamefont {Kittel}(1948)}]{KittelPR48}%
  \BibitemOpen
  \bibfield  {author} {\bibinfo {author} {\bibfnamefont {C.}~\bibnamefont
  {Kittel}},\ }\href {https://link.aps.org/doi/10.1103/PhysRev.73.155}
  {\bibfield  {journal} {\bibinfo  {journal} {Phys. Rev.}\ }\textbf {\bibinfo
  {volume} {73}},\ \bibinfo {pages} {155} (\bibinfo {year} {1948})}\BibitemShut
  {NoStop}%
\bibitem [{\citenamefont {Maier-Flaig}\ \emph {et~al.}(2017)\citenamefont
  {Maier-Flaig}, \citenamefont {Klingler}, \citenamefont {Dubs}, \citenamefont
  {Surzhenko}, \citenamefont {Gross}, \citenamefont {Weiler}, \citenamefont
  {Huebl},\ and\ \citenamefont {Goennenwein}}]{MaierFlaigPRB17}%
  \BibitemOpen
  \bibfield  {author} {\bibinfo {author} {\bibfnamefont {H.}~\bibnamefont
  {Maier-Flaig}}, \bibinfo {author} {\bibfnamefont {S.}~\bibnamefont
  {Klingler}}, \bibinfo {author} {\bibfnamefont {C.}~\bibnamefont {Dubs}},
  \bibinfo {author} {\bibfnamefont {O.}~\bibnamefont {Surzhenko}}, \bibinfo
  {author} {\bibfnamefont {R.}~\bibnamefont {Gross}}, \bibinfo {author}
  {\bibfnamefont {M.}~\bibnamefont {Weiler}}, \bibinfo {author} {\bibfnamefont
  {H.}~\bibnamefont {Huebl}}, \ and\ \bibinfo {author} {\bibfnamefont
  {S.~T.~B.}\ \bibnamefont {Goennenwein}},\ }\href {\doibase
  10.1103/PhysRevB.95.214423} {\bibfield  {journal} {\bibinfo  {journal} {Phys.
  Rev. B}\ }\textbf {\bibinfo {volume} {95}},\ \bibinfo {pages} {214423}
  (\bibinfo {year} {2017})}\BibitemShut {NoStop}%
\bibitem [{\citenamefont {Maksymov}\ and\ \citenamefont
  {Kostylev}(2015)}]{MaksymovPhysE15}%
  \BibitemOpen
  \bibfield  {author} {\bibinfo {author} {\bibfnamefont {I.~S.}\ \bibnamefont
  {Maksymov}}\ and\ \bibinfo {author} {\bibfnamefont {M.}~\bibnamefont
  {Kostylev}},\ }\href {\doibase https://doi.org/10.1016/j.physe.2014.12.027}
  {\bibfield  {journal} {\bibinfo  {journal} {Physica E: Low-dimensional
  Systems and Nanostructures}\ }\textbf {\bibinfo {volume} {69}},\ \bibinfo
  {pages} {253} (\bibinfo {year} {2015})}\BibitemShut {NoStop}%
\bibitem [{\citenamefont {Kalinikos}\ and\ \citenamefont
  {Slavin}(1986)}]{KalinikosJPhysC86}%
  \BibitemOpen
  \bibfield  {author} {\bibinfo {author} {\bibfnamefont {B.~A.}\ \bibnamefont
  {Kalinikos}}\ and\ \bibinfo {author} {\bibfnamefont {A.~N.}\ \bibnamefont
  {Slavin}},\ }\href {\doibase 10.1088/0022-3719/19/35/014} {\bibfield
  {journal} {\bibinfo  {journal} {Journal of Physics C: Solid State Physics}\
  }\textbf {\bibinfo {volume} {19}},\ \bibinfo {pages} {7013} (\bibinfo {year}
  {1986})}\BibitemShut {NoStop}%
\bibitem [{\citenamefont {Demokritov}\ and\ \citenamefont
  {Slavin}(2021)}]{DemokritovSpringer21}%
  \BibitemOpen
  \bibfield  {author} {\bibinfo {author} {\bibfnamefont {S.~O.}\ \bibnamefont
  {Demokritov}}\ and\ \bibinfo {author} {\bibfnamefont {A.~N.}\ \bibnamefont
  {Slavin}},\ }\enquote {\bibinfo {title} {Spin waves},}\ in\ \href {\doibase
  10.1007/978-3-030-63210-6_6} {\emph {\bibinfo {booktitle} {Handbook of
  Magnetism and Magnetic Materials}}},\ \bibinfo {editor} {edited by\ \bibinfo
  {editor} {\bibfnamefont {J.~M.~D.}\ \bibnamefont {Coey}}\ and\ \bibinfo
  {editor} {\bibfnamefont {S.~S.}\ \bibnamefont {Parkin}}}\ (\bibinfo
  {publisher} {Springer International Publishing},\ \bibinfo {address} {Cham},\
  \bibinfo {year} {2021})\ pp.\ \bibinfo {pages} {281--346}\BibitemShut
  {NoStop}%
\bibitem [{\citenamefont {Golovchanskiy}\ \emph {et~al.}(2023)\citenamefont
  {Golovchanskiy}, \citenamefont {Abramov}, \citenamefont {Emelyanova},
  \citenamefont {Shchetinin}, \citenamefont {Ryazanov}, \citenamefont
  {Golubov},\ and\ \citenamefont {Stolyarov}}]{GolovchanskiyPRAppl23}%
  \BibitemOpen
  \bibfield  {author} {\bibinfo {author} {\bibfnamefont {I.}~\bibnamefont
  {Golovchanskiy}}, \bibinfo {author} {\bibfnamefont {N.}~\bibnamefont
  {Abramov}}, \bibinfo {author} {\bibfnamefont {O.}~\bibnamefont {Emelyanova}},
  \bibinfo {author} {\bibfnamefont {I.}~\bibnamefont {Shchetinin}}, \bibinfo
  {author} {\bibfnamefont {V.}~\bibnamefont {Ryazanov}}, \bibinfo {author}
  {\bibfnamefont {A.}~\bibnamefont {Golubov}}, \ and\ \bibinfo {author}
  {\bibfnamefont {V.}~\bibnamefont {Stolyarov}},\ }\href {\doibase
  10.1103/PhysRevApplied.19.034025} {\bibfield  {journal} {\bibinfo  {journal}
  {Phys. Rev. Appl.}\ }\textbf {\bibinfo {volume} {19}},\ \bibinfo {pages}
  {034025} (\bibinfo {year} {2023})}\BibitemShut {NoStop}%
\bibitem [{\citenamefont {Tosi}\ \emph {et~al.}(2014)\citenamefont {Tosi},
  \citenamefont {Mohiyaddin}, \citenamefont {Huebl},\ and\ \citenamefont
  {Morello}}]{TosiAIPAdv14}%
  \BibitemOpen
  \bibfield  {author} {\bibinfo {author} {\bibfnamefont {G.}~\bibnamefont
  {Tosi}}, \bibinfo {author} {\bibfnamefont {F.~A.}\ \bibnamefont
  {Mohiyaddin}}, \bibinfo {author} {\bibfnamefont {H.}~\bibnamefont {Huebl}}, \
  and\ \bibinfo {author} {\bibfnamefont {A.}~\bibnamefont {Morello}},\ }\href
  {\doibase 10.1063/1.4893242} {\bibfield  {journal} {\bibinfo  {journal} {AIP
  Advances}\ }\textbf {\bibinfo {volume} {4}},\ \bibinfo {pages} {087122}
  (\bibinfo {year} {2014})}\BibitemShut {NoStop}%
\bibitem [{\citenamefont {Krupka}\ \emph {et~al.}(2013)\citenamefont {Krupka},
  \citenamefont {Wosik}, \citenamefont {Jastrzebski}, \citenamefont {Ciuk},
  \citenamefont {Mazierska},\ and\ \citenamefont {Zdrojek}}]{KrupkaIEEE13}%
  \BibitemOpen
  \bibfield  {author} {\bibinfo {author} {\bibfnamefont {J.}~\bibnamefont
  {Krupka}}, \bibinfo {author} {\bibfnamefont {J.}~\bibnamefont {Wosik}},
  \bibinfo {author} {\bibfnamefont {C.}~\bibnamefont {Jastrzebski}}, \bibinfo
  {author} {\bibfnamefont {T.}~\bibnamefont {Ciuk}}, \bibinfo {author}
  {\bibfnamefont {J.}~\bibnamefont {Mazierska}}, \ and\ \bibinfo {author}
  {\bibfnamefont {M.}~\bibnamefont {Zdrojek}},\ }\href {\doibase
  10.1109/TASC.2012.2237515} {\bibfield  {journal} {\bibinfo  {journal} {IEEE
  Transactions on Applied Superconductivity}\ }\textbf {\bibinfo {volume}
  {23}},\ \bibinfo {pages} {1501011} (\bibinfo {year} {2013})}\BibitemShut
  {NoStop}%
\bibitem [{\citenamefont {Solt}(2004)}]{SoltJAP62}%
  \BibitemOpen
  \bibfield  {author} {\bibinfo {author} {\bibfnamefont {J.}~\bibnamefont
  {Solt}, \bibfnamefont {Irvin~H.}},\ }\href {\doibase 10.1063/1.1728651}
  {\bibfield  {journal} {\bibinfo  {journal} {Journal of Applied Physics}\
  }\textbf {\bibinfo {volume} {33}},\ \bibinfo {pages} {1189} (\bibinfo {year}
  {2004})}\BibitemShut {NoStop}%
\bibitem [{\citenamefont {Lancaster}(1997)}]{Lancaster97}%
  \BibitemOpen
  \bibfield  {author} {\bibinfo {author} {\bibfnamefont {M.~J.}\ \bibnamefont
  {Lancaster}},\ }\href {\doibase 10.1017/CBO9780511526688} {\emph {\bibinfo
  {title} {Passive Microwave Device Applications of High-Temperature
  Superconductors}}}\ (\bibinfo  {publisher} {Cambridge University Press},\
  \bibinfo {year} {1997})\BibitemShut {NoStop}%
\bibitem [{\citenamefont {Klingler}\ \emph {et~al.}(2014)\citenamefont
  {Klingler}, \citenamefont {Chumak}, \citenamefont {Mewes}, \citenamefont
  {Khodadadi}, \citenamefont {Mewes}, \citenamefont {Dubs}, \citenamefont
  {Surzhenko}, \citenamefont {Hillebrands},\ and\ \citenamefont
  {Conca}}]{Klingler2014}%
  \BibitemOpen
  \bibfield  {author} {\bibinfo {author} {\bibfnamefont {S.}~\bibnamefont
  {Klingler}}, \bibinfo {author} {\bibfnamefont {A.~V.}\ \bibnamefont
  {Chumak}}, \bibinfo {author} {\bibfnamefont {T.}~\bibnamefont {Mewes}},
  \bibinfo {author} {\bibfnamefont {B.}~\bibnamefont {Khodadadi}}, \bibinfo
  {author} {\bibfnamefont {C.}~\bibnamefont {Mewes}}, \bibinfo {author}
  {\bibfnamefont {C.}~\bibnamefont {Dubs}}, \bibinfo {author} {\bibfnamefont
  {O.}~\bibnamefont {Surzhenko}}, \bibinfo {author} {\bibfnamefont
  {B.}~\bibnamefont {Hillebrands}}, \ and\ \bibinfo {author} {\bibfnamefont
  {A.}~\bibnamefont {Conca}},\ }\href {\doibase 10.1088/0022-3727/48/1/015001}
  {\bibfield  {journal} {\bibinfo  {journal} {Journal of Physics D: Applied
  Physics}\ }\textbf {\bibinfo {volume} {48}},\ \bibinfo {pages} {015001}
  (\bibinfo {year} {2014})}\BibitemShut {NoStop}%
\bibitem [{\citenamefont {Zyuzin}\ and\ \citenamefont
  {Bazhanov}(1996)}]{ZyuzinJETPLett96}%
  \BibitemOpen
  \bibfield  {author} {\bibinfo {author} {\bibfnamefont {A.~M.}\ \bibnamefont
  {Zyuzin}}\ and\ \bibinfo {author} {\bibfnamefont {A.~G.}\ \bibnamefont
  {Bazhanov}},\ }\href {\doibase 10.1134/1.567056} {\bibfield  {journal}
  {\bibinfo  {journal} {Journal of Experimental and Theoretical Physics
  Letters}\ }\textbf {\bibinfo {volume} {63}},\ \bibinfo {pages} {555}
  (\bibinfo {year} {1996})}\BibitemShut {NoStop}%
\bibitem [{\citenamefont {Cao}\ \emph {et~al.}(2015)\citenamefont {Cao},
  \citenamefont {Yan}, \citenamefont {Huebl}, \citenamefont {Goennenwein},\
  and\ \citenamefont {Bauer}}]{CaoPRB15}%
  \BibitemOpen
  \bibfield  {author} {\bibinfo {author} {\bibfnamefont {Y.}~\bibnamefont
  {Cao}}, \bibinfo {author} {\bibfnamefont {P.}~\bibnamefont {Yan}}, \bibinfo
  {author} {\bibfnamefont {H.}~\bibnamefont {Huebl}}, \bibinfo {author}
  {\bibfnamefont {S.~T.~B.}\ \bibnamefont {Goennenwein}}, \ and\ \bibinfo
  {author} {\bibfnamefont {G.~E.~W.}\ \bibnamefont {Bauer}},\ }\href {\doibase
  10.1103/PhysRevB.91.094423} {\bibfield  {journal} {\bibinfo  {journal} {Phys.
  Rev. B}\ }\textbf {\bibinfo {volume} {91}},\ \bibinfo {pages} {094423}
  (\bibinfo {year} {2015})}\BibitemShut {NoStop}%
\bibitem [{\citenamefont {Kittel}(1958)}]{KittelPR58}%
  \BibitemOpen
  \bibfield  {author} {\bibinfo {author} {\bibfnamefont {C.}~\bibnamefont
  {Kittel}},\ }\href {\doibase 10.1103/PhysRev.110.1295} {\bibfield  {journal}
  {\bibinfo  {journal} {Phys. Rev.}\ }\textbf {\bibinfo {volume} {110}},\
  \bibinfo {pages} {1295} (\bibinfo {year} {1958})}\BibitemShut {NoStop}%
\end{thebibliography}%

\clearpage
\newpage
\onecolumngrid
\renewcommand\thefigure{\thesection S\arabic{figure}} 
\setcounter{figure}{0}
\setcounter{equation}{0}

\section*{Supplemental material}

\subsection{Experimental methods}
\label{app:methods}

Coplanar waveguide (CPW) resonators and transmission lines were fabricated by optical lithography and etching with Ar plasma in a reactive ion etching (RIE) chamber starting from Au~(200~nm)/YBCO~(330~nm)/sapphire~(430~$\mu$m) films produced by Ceraco GmbH (M-type) and diced into $8 \times 5~\mathrm{mm}^2$ chips. The YBCO surface was exposed to oxygen plasma as a final step of the fabrication procedure. Gold pads defined at the edge of the chip were used to facilitate the bonding to an external printed circuit board. 

Transmission spectra were acquired as a function of the microwave frequency by means of a vector network analyzer, which is connected to the printed circuit board by coaxial cables whose insertion loss was removed from transmission data. The estimated incident power considering line attenuation is $P_{inc} \approx -15$~dBm. The YIG/GGG film was placed on the YBCO surface and kept in position thanks to a plastic screw that gently pushed the GGG substrate from the backside \cite{GhirriPrAppl23}. The experiments were carried out in a variable temperature cryostat equipped with a superconducting solenoid for the generation of the static magnetic field.

\subsection{Temperature dependence of the saturation magnetization}
\label{app:Msat}

The saturation magnetization of YIG ($M_s$) is expected to be weakly dependent on temperature in the 10-90~K range. Following Refs.~\cite{SoltJAP62, MaierFlaigPRB17}, we assume that its temperature behavior is given by 
\begin{equation}
    M_s=M_0(1-u T^{3/2} - v T^{5/2}),
    \label{eq:suppl:Ms}
\end{equation}
where $u=23 \times 10^{-6}~\mathrm{K^{-3/2}}$ and $v=1.08 \times 10^{-7}~\mathrm{K^{-5/2}}$ \cite{MaierFlaigPRB17}. In our case we used $\mu_0 M_0=0.253$~T.

\subsection{Modelization of the CPW resonator}
\label{app:LC_model}

To transmission spectrum of the CPW resonators can be reproduced with \cite{Hein}
\begin{equation}
S_{21}=\frac{10^{-IL/20}}{\sqrt{1+Q_L^2(\omega/\omega_c- \omega_c/\omega)^2}},
\label{suppl:lorentzian}
\end{equation}
where $\omega_c$ is the fundamental mode frequency, $Q_L$ is the loaded quality factor and $IL$ is the insertion loss. The values obtained at 10~K are $\omega_c/2\pi=10.097$~GHz, $Q_L=3628.4$ and $IL=19.768$~dB (Inset in Fig.~1(c) of the Letter).

Following Ref.~\cite{GhigoSuperSciTech04}, the temperature dependence of the frequency of the resonator ($\omega_c$) can be evaluated as
\begin{eqnarray}
   \omega_c(T) = \omega_c(0) \sqrt{\frac{L(T_0)C(T_0)}{L(T)C(T)}},
\end{eqnarray}
where $\omega_c(0)=\omega_c(T_0)$ and $T_0=10$~K. Here $L$ and $C$ are respectively the inductance and the capacitance per unit length. By neglecting the weak temperature variations of the latter \cite{GhigoSuperSciTech04}, we obtain
\begin{eqnarray}
   \omega_c(T) \approx \omega_c(0) \sqrt{\frac{L(T_0)}{L(T)}}=\omega_c(0) \sqrt{\frac{L_g+L_k[\lambda_L (T_0)]}{L_g+L_k[\lambda_L (T)]}},
   \label{suppl:eq:omega0-vs-temp}
\end{eqnarray}
being $L_g$ and $L_k$ the geometric and kinetic inductance per unit length, respectively. The geometric inductance is temperature independent and reads \cite{Lancaster97}
\begin{equation}
L_g=\mu_0 \frac{K(\sqrt{1-(w/a)^2})}{4 K(w/a)}
\end{equation}
where $K$ is the complete elliptic integral and $a=w+2s$. Conversely, the kinetic inductance depends on the temperature through the penetration depth (Eq.~2 in the Letter). For a CPW line, $L_k$ can be written as \cite{VendikIEEE98}
\begin{equation}
L_k=\frac{\mu_0}{d w_{eff}(\lambda_L)}\lambda_L^2,
\end{equation}
where 
\begin{equation}
\frac{1}{w_{eff}}=\frac{1}{w}(C_1+C_2).
\end{equation}
The expressions for the coefficients $C_1$ and $C_2$ are 
\begin{align}
C_1 = &\frac{2}{[K(\frac{w}{a})]^2}\frac{1}{1-(\frac{w}{a})^2} \\
& \left [ 1+\frac{1}{4} \ln (\frac{w}{\lambda_{L,\perp}}-1)- \frac{w}{4 a} \ln(\frac{a+w-2\lambda_{L,\perp}}{a-w+2\lambda_{L,\perp}}) \right ]
\end{align}
and
\begin{align}
C_2= & \frac{2}{[K(\frac{w}{a})]^2}\frac{\frac{w}{a}}{1-(\frac{w}{a})^2} \\
&\left [ 1+\frac{1}{4} \ln (\frac{a}{\lambda_{L,\perp}}+1)- \frac{a}{4 w} \ln(\frac{a+w-2\lambda_{L,\perp}}{a-w+2\lambda_{L,\perp}}) \right ],
\end{align}
being $\lambda_{L,\perp}=2 \lambda_L^2/t$ \cite{GhigoSuperSciTech04}.

\subsection{Dissipation and decay of the magnetic field in the superconducting layer}
\label{app:decay_kappa}

According to the two-fluid model, the conductivity of the superconductor is $\sigma (\omega)=\sigma_1(\omega)-i \sigma_2(\omega)$, where $\sigma_1$ is responsible for the dissipation and fully describes the conductivity above $T_c$, while $\sigma_2$ is responsible for the kinetic inductance. In the normal state the skin depth of the electromagnetic field can be defined as $\delta_s=\sqrt{2 / \omega \mu_0 \sigma_1}$. 

Data reported in ref.~\cite{KrupkaIEEE13} for 280~nm thick YBCO/sapphire films at 28.2~GHz, indicate that $\sigma_1$ is in the range between $0.5 \times 10^6$ and $1.2 \times 10^6$~S/m, while $\sigma_2$ rapidly increases from $\approx 4 \times 10^6$~S/m at $T\approx T_c$ to saturate at $\approx 4 \times 10^7$~S/m at 13~K. The ratio $\sigma_1/\sigma_2$ results between 0.33 and 0.12 from $T_c$ to 83~K, and lower than 0.1 below 83~K. In our experiment, we expect even lower $\sigma_1/\sigma_2$ ratios due to the higher thickness of the film ($t=330$~nm) and to the lower frequency of the resonator, $\omega_c(T)$. We thus neglect $\sigma_1$ and assume that the complex conductivity is $\sigma=-i/\omega \mu_0 \lambda_L^2(T)$ as in the case of a perfect superconductor.

Within the superconducting layer, the decay constant of the magnetic field generated by the spin wave is \cite{BorstScience23}
\begin{equation}
\kappa= k \sqrt{1+ i \omega \mu_0 \sigma / k^2 } = k \sqrt{1+ \frac{2 i }{k^2 \delta_s(T)^2} \frac{n_n(T)}{n}+\frac{1}{k^2 \lambda_L^2(T)}},
\label{eq:suppl_kappa}
\end{equation}
where $n_n(T)/n$ is the fractional number density of normal electrons. The estimated skin depth results $\delta_s > 10~\mathrm{\mu m}$ at 10~GHz, thus much larger than both the penetration depth in the superconducting state $\lambda_L(T)$ (Fig.~1(c) of the Letter) and the thickness of the YBCO layer. 
By neglecting the second term under the square root, Eq.~\ref{eq:suppl_kappa} results $\kappa \approx \sqrt{k^2+1/\lambda_L^2(T)}$ thus, considering that in our experiment $k^2 \lambda_L^2(T) \ll 1$,  we finally obtain $\kappa \approx 1/\lambda_L(T)$. 

\begin{figure*}[t]
\centering
\includegraphics[width=\linewidth]{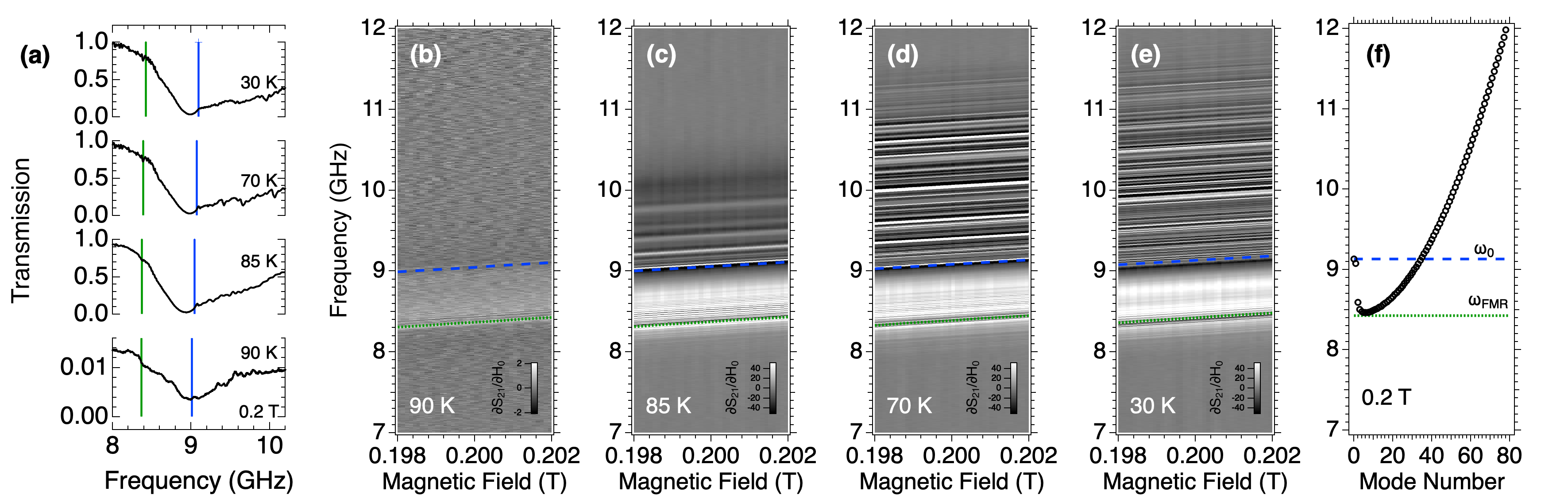}
\caption{(a) Temperature evolution of the $S_{21}$ spectra of the YIG film acquired by using the YBCO CPW broadband line ($H_0=0.2$~T). (b-e) Spectral maps acquired at different temperatures. (f) Frequency of the spin wave resonance modes calculated with Eq.~\ref{suppl:KS-dipersion} as a function of the mode number $j$.}
\label{fig:suppl:broadband}
\end{figure*}

\subsection{Higher spin wave resonance modes}
\label{suppl:sec:broadband}
The spectral maps in Fig.~\ref{fig:suppl:broadband}(a-e) display a series of the spectra of the YIG film acquired by means of the YBCO broadband CPW line with the static magnetic field applied perpendicular to the plane containing the oscillating magnetic field as in Fig.~1 of the Letter (Damon-Eshbach geometry). The spectral maps are characterized by a transmission dip with the presence of several closely spaced lines. At 90~K the spin wave resonance modes are approximately found between $\omega_0$ and $\omega_{FMR}$; at 85~K and below many additional modes appear at frequency also above $\omega_0$. We attribute this evolution, which is marked by clear a contrast between the spectra taken above and below $T_c$, to the effect of the superconductor.

We compare these data to the the predictions of the analytic model by Kalinikos and Slavin \cite{KalinikosJPhysC86}, which describes the dispersion of spin wave resonance (SWR) modes in a magnetic film by including the effects of both dipolar and exchange interactions. Although the effects of the superconductor are not included in this theory, which therefore does not allow the description of the interplay between superconducting and magnetic layer, we can used it to calculate the expected frequencies of the spin wave resonance modes. The lowest ($j=0$) mode displays a quasi-uniform spin wave amplitude in the thickness of the film; its frequency mostly derives from the dipolar energy and follows Eq.~4 of the Letter, which can be used to fit the experimental spectra and derive the wavenumber $k_y=3 \times 10^5~\mathrm{rad/m}$ (Fig. 2 of the Letter). Higher ($j > 0$) excitations, usually named Perpendicular Standing Spin Wave (PSSW) modes, are determined also by the effect of the exchange interaction and characterized by a finite number of modes for which the spin wave amplitude periodically varies along
the film thickness. Assuming unpinned surface spins, the perpendicular wavenumber is $k_z=j\pi/d$. The PSSW mode frequencies thus follows \cite{KalinikosJPhysC86} 

\begin{equation}
\omega_j = \gamma \mu_0 \sqrt{\left(H_0+H_{ex}^{(j)}\right)\left(H_0+H_{ex}^{(j)}+M_s F_{jj}\right)},
\label{suppl:Herring-Kittel}
\end{equation}
where $H_{ex}^{(j)}=\frac{2A}{M_s}\left(k_y^2+\left(\frac{j\pi}{d}\right)^2\right)$ is the exchange field and $F_{jj}$ is the matrix element of the dipole-dipole interaction. Eq.~\ref{suppl:Herring-Kittel} can be approximated as \cite{DemokritovSpringer21}
\begin{equation}
\omega_j = \gamma \mu_0 \sqrt{\left[H_0+\frac{2A}{M_s}\left(k_y^2+\left(\frac{j\pi}{d}\right)^2\right)\right]\left[H_0+\frac{2A}{M_s}\left(k_y^2+\left(\frac{j\pi}{d}\right)^2\right) +M_s+H_0 \left(\frac{Ms/H_0}{j \pi/d}\right)^2 k_y^2\right]},
\label{suppl:KS-dipersion}
\end{equation}
being $d$ the thickness of the YIG film and $A$ the exchange constant \cite{Klingler2014}.

We use Eq.~\ref{suppl:KS-dipersion} to calculate the spin wave mode spectrum by including realistic experimental parameters. We consider the nominal thickness of the YIG film ($d=5~\mathrm{\mu m}$) and estimate the exchange constant in the temperature range of interest as $A \approx 5~\mathrm{pJ/m}$ \cite{Klingler2014, ZyuzinJETPLett96}. Fig.~\ref{fig:suppl:broadband}(f) shows the PSSW mode frequencies calculated at 0.2~T as a function of the mode number ($j$). Due both to the uncertainty in the value of the exchange constant at low temperature \cite{ZyuzinJETPLett96} and to the difficulty in precisely indexing the large number of SWR modes visible in the spectrum, a quantitative comparison between the experiment and theoretical model is not possible. However, we note that the calculated frequency separation between subsequent modes (panel (f)) is comparable to the pitch observed in the experimental spectra (panels (a-e)). As the mode number increases, the frequency initially jumps from $\omega_0$ down to $\omega_{FMR}$, then it progressively increases to values much higher than $\omega_0$. On the basis of this analysis, we attribute the resonances observed in the experimental map at 90~K (panel (b)) to the lowest PSSW modes, while those found in the spectra at lower temperatures (panels (c-e)) to higher number PSSW modes, whose progressive appearance as temperature drops is related to the effects of the superconductor.

\subsection{Additional transmission spectra}
\label{suppl:sec:suppl_res}

Transmission spectra were acquired with a YIG/GGG film having area $\approx 3.8 \times 2~\mathrm{mm}^2$, which was positioned in the middle of the YBCO resonator. We collected two datasets in different experimental runs; the measured spectral maps are shown in Fig.~\ref{fig:suppl:data} (dataset A) and in Fig.~3 of the Letter (dataset B). The two datasets show equivalent features, with minor differences in the absolute value of frequencies and coupling strengths that can be attributed to the slightly different positioning of the YIG film in the two experiments.

\subsection{Effects of the higher microwave mode}

In this section we analyze the spectral features related to the coupling between spin waves and a higher microwave mode. In order to understand these effects, we first analyze the experimental results, we then discuss the outcome of finite-element electromagnetic simulations and finally introduce a model to describe the coupled spin wave - microwave modes.

\begin{figure*}[t]
\centering
\includegraphics[width=\linewidth]{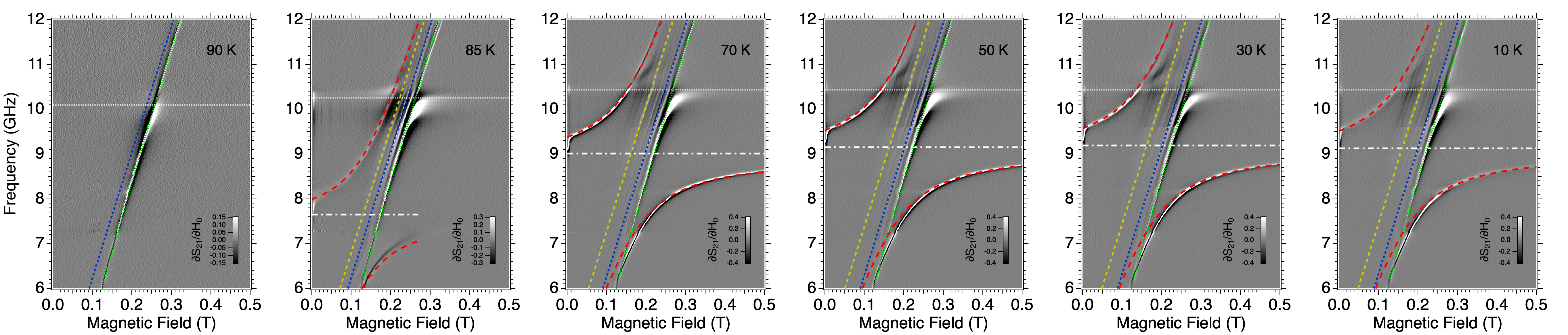}
\caption{Transmission spectral maps showing the evolution of the coupled spin wave and resonator modes at different temperatures (dataset A). The grey scale shows the numerical derivative of the transmission with respect to the magnetic field, $\partial S_{21}/\partial H_0$. The white dash-dot lines indicate the frequency of the resonator at each temperature, while white dotted lines the estimated frequency of the higher microwave mode. The dashed lines show $\omega_{FMR}(H_0)$ (green), $\omega_{0}(H_0)$ (blue), $\omega_{b}(H_0)$ (yellow) and the calculated polaritonic modes $\Omega_{\pm}(H_0)$ (red).}
\label{fig:suppl:data}
\end{figure*}

\begin{figure*}[t]
\centering
\includegraphics[width=12cm]{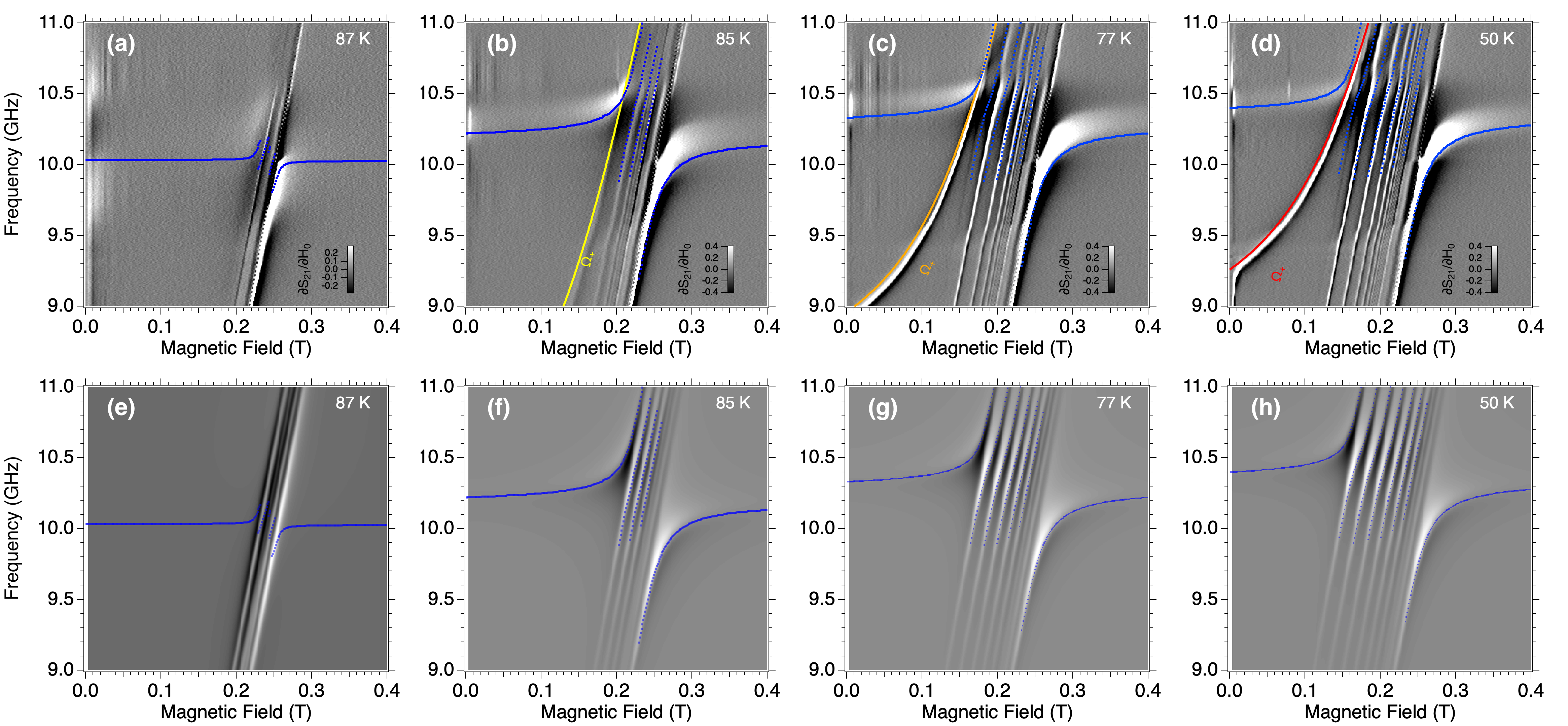}
\caption{Comparison between transmission spectral maps (a-d) measured at different temperatures and (e-d) calculated with Eq.~\ref{suppl:eq:S21_Cao}. Blue dotted lines indicate the calculated maximum transmission. Solid lines display the magnetic field dependence of the upper polariton ($\Omega_+$), as discussed in the Letter.
}
\label{fig:suppl:dielectric_mode}
\end{figure*}

\paragraph*{Transmission spectroscopy data.} Fig.~\ref{fig:suppl:dielectric_mode} shows a zoom  on the spectral maps reported in Fig.~3 of the Letter. The spectra taken in zero applied magnetic field ($H_0=0$) show the presence of a higher microwave mode having frequency of 10.03~GHz at 87~K and 10.35~GHz at 50~K (Fig.~\ref{fig:suppl:dielectric_mode}). This mode is visible also in the spectra of  Fig.~\ref{fig:suppl:data} and in those acquired with the supplemental resonator (Sect.~\ref{suppl:sec:suppl_res}). With respect to the fundamental mode of the YBCO CPW resonator (Fig.~1 of the Letter), the frequency of this mode has a weaker dependence from temperature; this discrepancy suggests a different distribution of the microwave field in the two cases.  


In the presence of an external magnetic field, the spectra evidence the coupling between the higher microwave mode and spin waves. In Fig.~\ref{fig:suppl:dielectric_mode}(a-d) we note the presence of multiple anticrossings, which are characterized by a broad polaritonic branch that converges towards $\omega_{FMR}$ for $H_0 \approx 0.25$~T and by several satellite lines at lower fields. This behavior closely resembles what obtained by placing the YIG film on a metal microstrip \cite{GhirriPrAppl23} or for ferromagnetic films in a cavity \cite{CaoPRB15}. However, the maps in Fig.~\ref{fig:suppl:dielectric_mode}(a-d) show a number of satellite lines that increases as the main anticrossing related to the fundamental CPW mode gets wider, i.e.~as the temperature decreases. We note that at each temperature the position of the upper polariton ($\Omega_+$) sets the lower limit of the multiple anticrossings in Fig.~\ref{fig:suppl:dielectric_mode}. 

\begin{figure}[t]
\centering
\includegraphics[width=11cm]{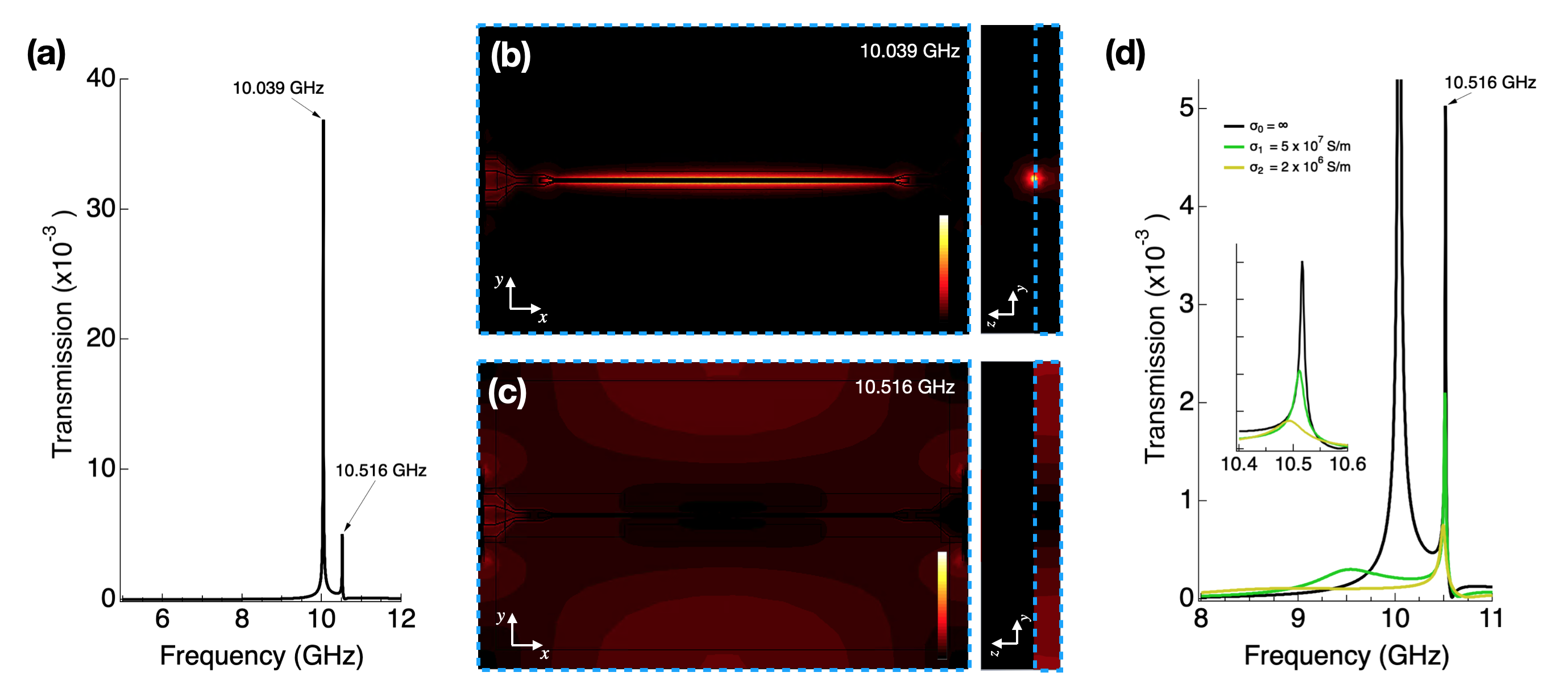}
\caption{Finite-element electromagnetic simulation of the bare resonator. (a) Transmission spectrum calculated between 5 and 12~GHz by approximating the YBCO film as a perfect conductor. (b,c) Distribution of the amplitude of $\mathbf{h_{ac}}$ calculated respectively for the modes at 10.039 and 10.516~GHz. For both frequencies, the maps on the left show the field calculated in the $z=0$ plane; conversely the smaller map on the right shows the field in the $x=0$ plane. Cyan dashed rectangles delimit the area of the sapphire substrate. The color scale is the same in all plots. (d) Evolution of the transmission spectrum for different values of the film conductivity $\sigma$.}
\label{fig:suppl:simulation}
\end{figure}

\paragraph*{Electromagnetic simulations.} Finite-element electromagnetic simulations (CST-Microwave Studio) were carried out for the bare resonator in a wide frequency range between 5 and 12~GHz (Fig.~\ref{fig:suppl:simulation}). The resonator was modelled with realistic geometry and materials as discussed in Ref.~\cite{GhirriPrAppl23}. The superconductor was initially considered as a perfect electric conductor having infinite dc conductivity ($\sigma_0 = \infty$). The transmission ($S_{21}$) spectrum obtained from the simulation is shown in panel (a). The fundamental mode of the CPW resonator is found at 10.039~GHz, in good agreement with the experimental data (Fig.~1 of the Letter). The distribution of the oscillating magnetic field ($\mathbf{h_{ac}}$) shows an antinode in the middle of the CPW resonator with the field concentrated around the central conductor (Fig.~\ref{fig:suppl:simulation}(b)), as expected. At 10.516~GHz an additional mode appears with lower amplitude respect to the fundamental one (panel (a)). The distribution of $\mathbf{h_{ac}}$ calculated for this mode (panel (c)) shows that the magnetic field is nonuniformly spread across the whole chip, in particular into the sapphire substrate, while the amplitude of $\mathbf{h_{ac}}$ is minimum in the middle of the chip. We thus hereafter refer to the latter mode as the ``dielectric'' mode. 

Simulations carried out by considering a layer of finite conductivity ($\sigma_{1,2} \ne \infty$) in place of the perfect electric conductor, show that the amplitude and frequency of the fundamental mode drop as the conductivity of the film decreases (Fig.~\ref{fig:suppl:simulation}(d)). The same variation of $\sigma$ determines less pronounced changes of both amplitude and frequency of the dielectric mode. These trends are in line with those evidenced by transmission spectra taken at different temperatures across the $T_c$ of YBCO (Fig.~3 of the Letter and Fig.~\ref{fig:suppl:data}).

\paragraph*{Model.} The model by Cao et al.~in Ref.~\cite{CaoPRB15} describes the coupling between a microwave cavity having frequency $\omega_c$ and the spin wave resonance (SWR) modes in a magnetic film of thickness $d$. The transmission coefficient follows
\begin{equation}
S_{21}= \frac{\kappa_c}{i (\omega-\omega_c)-\kappa_c - i \sum_j \frac{g^2}{\omega - \omega_{SWR}^{(j)}+ i \kappa_S}},
\label{suppl:eq:S21_Cao}
\end{equation}
where $\kappa_c$ and $\kappa_s$ are respectively the cavity and spin damping rates and $g$ is the collective coupling strength between the cavity mode and the $j$-th SWR mode. 

\begin{table}[h]
    \centering
    \begin{tabular}{ccccc}
        \hline
        \hline
        $T$(K) & $\omega_{c}$(GHz) & $\tilde{d}(\mathrm{\mu m})$ & $n_{SWR}$ & $g$(GHz) \\
        \hline
        87 & 10.030 & 1.40 & 12 & 0.04 \\
        85 & 10.190 & 1.40 & 16 & 0.18 \\
        77 & 10.289 & 1.35 & 20 & 0.18 \\
        50 & 10.350 & 1.35 & 22 & 0.18 \\
        \hline
        \hline
    \end{tabular}
    \caption{List of the parameters used to fit the spectral maps in Fig.~\ref{fig:suppl:dielectric_mode}.}
    \label{suppl:tab:fit_dielectric_mode}
\end{table}

As pointed out by Kittel, resonances with symmetrically pinned boundaries (even mode numbers) can efficiently absorb energy and are observed with a strong amplitude, whereas resonances with odd mode numbers have smaller amplitude and are observed only when the microwave field is inhomogeneous \cite{KittelPR58}. We fitted the experimental data in Fig.~\ref{fig:suppl:dielectric_mode} by using Eq.~\ref{suppl:eq:S21_Cao}. We considered only SWR modes having even $j$, whose frequency has been obtained using Eq.~\ref{suppl:KS-dipersion}. We note that the energy separation between the SWR modes in Fig.~\ref{fig:suppl:dielectric_mode} is larger than what obtained from the calculated PSSW modes (Fig.~\ref{fig:suppl:broadband}). A satisfactory fit of the experimental spectra can be obtained by considering an effective thickness $\tilde{d}<d$ to account for the frequency separation between the modes. The fitted values of $\tilde{d}$ are in the $1.35-1.40~\mathrm{\mu m}$ range across the whole temperature range, however such changes may be caused also by variations of the the exchange constant as a function of the temperature. The maximum number of SWR modes ($n_{SWR}$) has been used as free parameter in order to match the number of lines observed in the experimental maps while we consider $g$ constant for each $j$. The obtained parameters are summarized in Table~\ref{suppl:tab:fit_dielectric_mode}; the calculated curves show a good correspondence with the experimental data (Fig.~\ref{fig:suppl:dielectric_mode}). In particular, the tails of these anticrossing are visible in a wide frequency range down to about 7~GHz; this behavior justifies the presence of additional resonances in the spectral maps shown in Fig.~4 of the Letter.\\

\begin{figure}[bh]
\centering
\includegraphics[width=8cm] {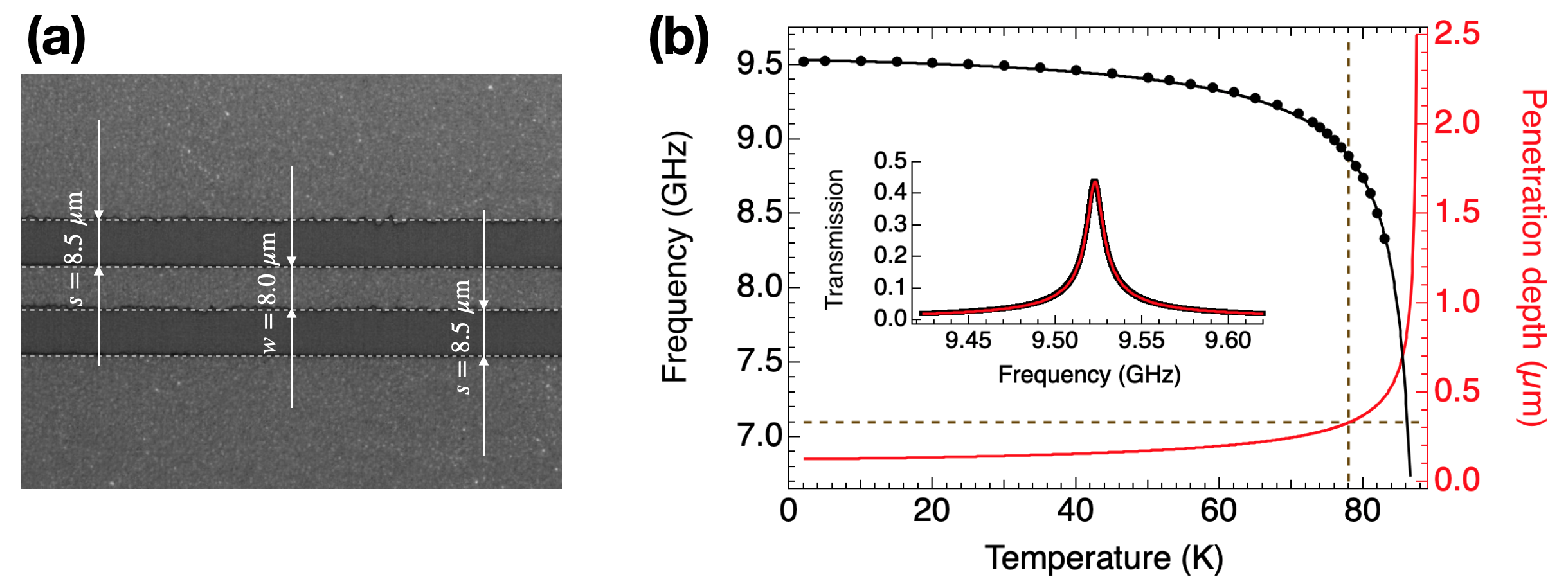}
\caption{Characterization of the bare supplemental resonator. (a) Image captured with the optical microscope showing the central region of the CPW resonator. The patterned YBCO film is light grey while the sapphire substrate is dark grey. (b)  Temperature dependence of the fundamental frequency of the bare resonator (circles); the black line is calculated with Eq.~\ref{suppl:eq:omega0-vs-temp}. The penetration depth derived at the frequency of the resonator is displayed in red. The horizontal dashed line indicates the thickness of the YBCO film ($t=330$~nm). Inset. Transmission spectrum taken at 10~K (circles); the red line shows the curve calculated with Eq.~\ref{suppl:lorentzian} and the fitting parameters $\omega_c/2\pi=9.523$~GHz, $Q_L=1146.5$ and $IL=7.11$~dB.}
\label{fig:suppl:narrower_bare}
\end{figure}

To summarize, in this section we have analyzed the coupling between the dielectric mode and SWR resonances in the YIG film. We point out that there are several factors that might affect the SWR spectrum \cite{ZyuzinJETPLett96}, such as the inhomogenous microwave field, that might justify the effective value of the film thickness used to fit the frequency separation between the resonances (Fig.~\ref{fig:suppl:dielectric_mode}). The spectra also show that the number of these lines is correlated to splitting of the main anticrossing, suggesting that these modes are probably excited as a consequence of the hybridization between CPW resonator and YIG film. However, we note in Fig.~\ref{fig:suppl:dielectric_mode} that the upper polariton crosses the dielectric mode at $\approx 10.5$~GHz without any noticeable repulsion between the lines. This behavior suggests that the simultaneous hybridization among YIG film, CPW resonator and dielectric mode is vanishingly small in our experiment. 


From the comparison between datasets in Fig.~\ref{fig:suppl:dielectric_mode}, Fig.~\ref{fig:suppl:data} and reported in Ref.~\cite{GhirriPrAppl23}, we note that the features of the additional anticrossings depend by the specific experimental conditions. In particular, in the latter two cases the additional lines are evident for frequencies around $\approx 10.5$~GHz but they become barely visible at the frequency of the CPW resonator. This observation gives further justification to the analysis reported above, in which the additional anticrossings are attributed to the coupling between dielectric mode and YIG film.

\subsection{Experiments with the supplemental resonator}
\label{app:additional_data}

\begin{figure*}[t]
\centering
\includegraphics[width=\linewidth]{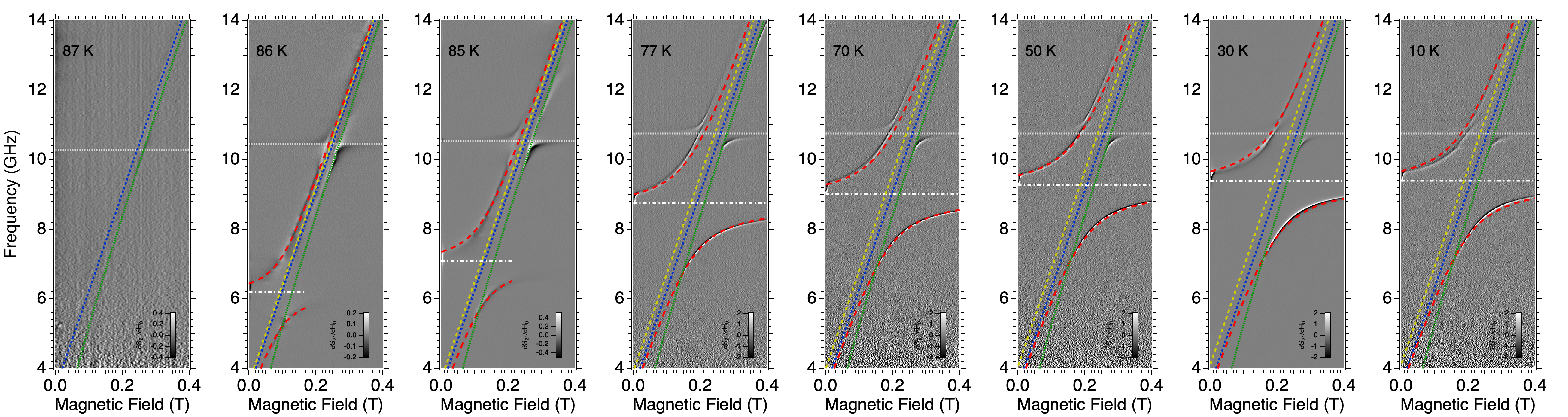}
\caption{Transmission spectra acquired with the the YIG/GGG film loaded onto the supplemental resonator. The grey scale shows the numerical derivative of the transmission with respect to the magnetic field, $\partial S_{21}/\partial H_0$. The dashed lines show $\omega_{FMR}(H_0)$ (green), $\omega_{0}(H_0)$ (blue), $\omega_{b}(H_0)$ (yellow) and the calculated polaritonic modes $\Omega_{\pm}(H_0)$ (red). White dash-dot lines show the frequency of the resonator at each temperature, while white dotted lines indicate the frequency of the dielectric mode.}
\label{fig:suppl:narrower_res}
\end{figure*}

To better assess the robustness of the results reported in the Letter, we fabricated an additional YBCO/sapphire CPW resonator having different geometric parameters. We tested its behavior by transmission spectroscopy measurements at different temperatures, firstly for the bare resonator and subsequently with the same YIG/GGG film as used with the main resonator.

\paragraph*{Characterization of the bare resonator.} The fabrication procedure described in Sect.~\ref{app:methods} and Ref.~\cite{GhirriPrAppl23} was applied to produce a YBCO/sapphire CPW resonator having a central conductor with mean width $w=8.0~\mathrm{\mu m}$ and separation $s=8.5~\mathrm{\mu m}$ between the central conductor and the lateral ground planes (Fig.~\ref{fig:suppl:narrower_bare}(a)). The estimated characteristic impedance is $Z_0=62~\mathrm{\Omega}$. The length of the central conductor of the resonator is 6.2~mm and the capacitive coupling gaps are $50~\mathrm{\mu m}$ wide.

The transmission spectrum acquired at 10~K shows a peak at 9.52~GHz that corresponds to the fundamental mode of the CPW resonator (Inset in Fig.~\ref{fig:suppl:narrower_bare}(b)). The evolution of the peak frequency was tracked as a function of temperature and fit with Eq.~\ref{suppl:eq:omega0-vs-temp} (Fig.~\ref{fig:suppl:narrower_bare}(b)). The parameters that characterize the temperature dependence of the penetration depth (Eq.~2 of the Letter) are $\lambda_L(0)=125~\mathrm{nm}$, $T_c=87.6~\mathrm{K}$ and $p=4/3$, in close agreement with those reported for the main resonator.

\paragraph*{Hybrid magnon-photon system.} The YIG/GGG film was positioned onto the resonator as described in Sect.~\ref{app:methods}. The experimental spectra in Fig.~\ref{fig:suppl:narrower_res} were acquired at different temperatures to monitor the evolution of the spectral maps. Overall, the temperature dependence of the normal mode spectrum is consistent with what obtained with the main resonator. In particular these data confirms, as temperature decreases, the progressive increase of the resonator frequency, the widening of the anticrossing gap and the shift of the upper polariton towards lower magnetic fields.

By using of the model described in the Letter, we calculated the dispersion of the polaritonic modes at each temperature. In order to reproduce the experimental spectra, we extracted from each map the frequency of the resonator ($\omega_c$), while the frequency shift ($\delta_{sc}$) and coupling strength ($g$) were calculated directly from Eqs.~6 and 7 of the Letter by including the temperature independent wavenumber, $k_y=4 \times 10^5~\mathrm{rad~m^{-1}}$, $r=0.047$ and number of spins, $N_s=1.8 \times 10^{14}$, as fitting parameters. The latter resulted about one fifth of the values reported for the main resonator. 

\begin{figure}[t]
\centering
\includegraphics[width=5cm]{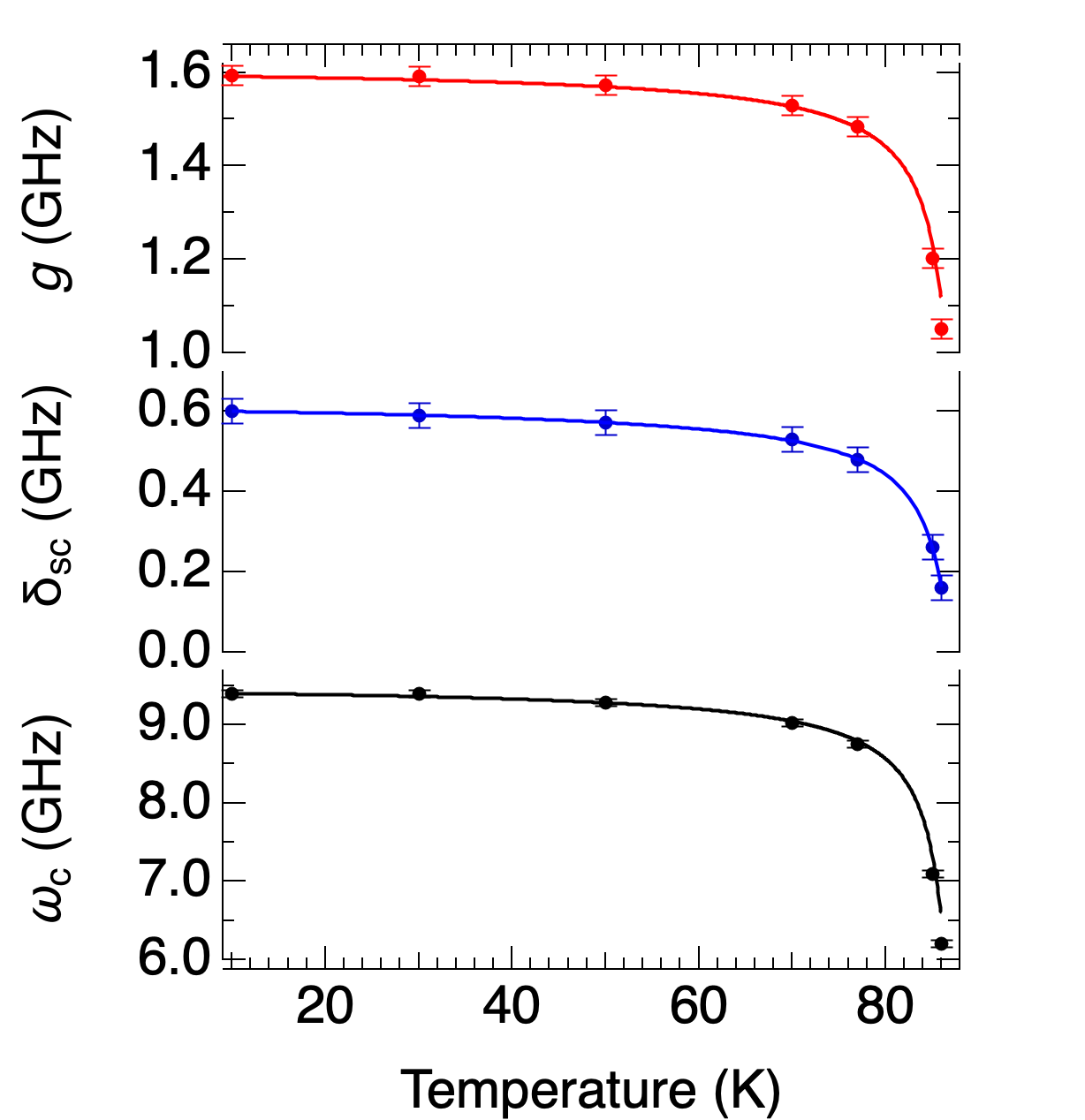}
\caption{Temperature dependence of the parameters used to reproduce the evolution of the spectra measured with the supplemental resonator (circles). Solid lines are calculated as discussed in the text.}
\label{fig:suppl:narrower_param}
\end{figure}

Fig.~\ref{fig:suppl:narrower_res} shows that at each temperature the calculated polaritonic modes compare well with the experimental spectra. The temperature evolution of the obtained $\omega_c$, $\delta_{sc}$ and $g$ parameters is reported in Fig.~\ref{fig:suppl:narrower_param}. In the presence of the YIG/GGG sample, the fit of the temperature dependence of $\omega_c$ (Eq.~\ref{suppl:eq:omega0-vs-temp}) provides a slightly reduced critical temperature ($T_c=86.9~\mathrm{K}$), in agreement to what observed with the main resonator. The maximum collective coupling strength, $g_0/2 \pi \approx 1.6~\mathrm{GHz}$, is slightly lower than those obtained with the main resonator. This value is determined by the lower volume of the supplemental resonator which, in spite of the larger spin-photon coupling (Eq.~7 of the Letter), gives rise to a lower $N_s$. The maximum frequency shift $\delta_{sc}/2 \pi \approx 0.6~\mathrm{GHz}$ resulted lower than what achieved with the main resonator, while its temperature dependence is consistent with the evolution of the polariton modes and the fitted $k_y$. 

In summary, the results obtained with the supplemental resonator support the validity of the analysis and of the main conclusions reported in the Letter.

\end{document}